\documentclass[preprint]{aastex}
\newcommand{\WMAP}{\textsl{WMAP}}

\newcommand{\Bennett}{{C. L. Bennett}}

\newcommand{\Halpern}{{M. Halpern}}

\newcommand{\Hinshaw}{{G. Hinshaw}}
\newcommand{\Jarosik}{{N. Jarosik}}
\newcommand{\Kogut}{{A. Kogut}}

\newcommand{\Limon}{{M. Limon}}
\newcommand{\Meyer}{{S. S. Meyer}}
\newcommand{\Nolta}{{M. R. Nolta}}

\newcommand{\Page}{{L. Page}}
\newcommand{\Peiris}{{H. V. Peiris}}
\newcommand{\Spergel}{{D. N. Spergel}}
\newcommand{\Tucker}{{G. S. Tucker}}
\newcommand{\Verde}{{L. Verde}}

\newcommand{\Wollack}{{E. Wollack}}
\newcommand{\Wright}{{E. L. Wright}}
\newcommand{\Brown}{{Dept. of Physics, Brown University, %
            Providence, RI 02912}}
\newcommand{\Goddard}{{Code 685, Goddard Space Flight Center, %
            Greenbelt, MD 20771}}
\newcommand{\NRCFellow}{{National Research Council (NRC) Fellow}}
\newcommand{\PrincetonPhysics}{{Dept. of Physics, Jadwin Hall, %
            Princeton University, Princeton, NJ 08544}}
\newcommand{\PrincetonAstro}{{Dept. of Astrophysical Sciences, %
            Princeton University, Princeton, NJ 08544}}

\newcommand{\UBC}{{Dept. of Physics and Astronomy, University of %
            British Columbia, Vancouver, BC  Canada V6T 1Z1}}
\newcommand{\UChicago}{{Depts. of Astrophysics and Physics, EFI and CfCP, %
            University of Chicago, Chicago, IL 60637}}
\newcommand{\UCLA}{{UCLA Astronomy, PO Box 951562, Los Angeles, CA 90095-1562}}
\shortauthors{Verde et al.}
\usepackage{graphicx}
\usepackage{natbib}
\usepackage{epsfig}
\usepackage{amssymb}

\newcommand{\lya}{Lyman $\alpha$ }

\newcommand{\ba}{\begin{eqnarray}}
\newcommand{\ea}{\end{eqnarray}}
\newcommand{\be}{\begin{equation}}
\newcommand{\ee}{\end{equation}} 
\newcommand{\C}{{\cal C}}

\newcommand{\cL}{{\cal L}}

\slugcomment{Accepted by the \emph{Astrophysical Journal}}
\begin{document}
\title{First Year {\sl Wilkinson Microwave Anisotropy Probe} ({\sl WMAP})\altaffilmark{1} Observations: Parameter
Estimation Methodology}
\author{
\Verde \altaffilmark{2,3}, 
\Peiris\altaffilmark{2},
\Spergel \altaffilmark{2}, 
\Nolta \altaffilmark{6},
\Bennett \altaffilmark{4},  
\Halpern \altaffilmark{5},
\Hinshaw\altaffilmark{4}, 
\Jarosik \altaffilmark{6}, \Kogut \altaffilmark{4},
\Limon \altaffilmark{4,8}, \Meyer \altaffilmark{7},
\Page \altaffilmark{6}, 
\Tucker \altaffilmark{4,8,9}, 
\Wollack \altaffilmark{4}, \Wright \altaffilmark{10}}

\altaffiltext{1}{{\sl WMAP} is the result of a partnership between Princeton University and NASA's Goddard Space Flight Center. Scientific guidance is provided by the {\sl WMAP} Science Team.}
\altaffiltext{2}{\PrincetonAstro}
\altaffiltext{3}{Chandra Fellow}
\altaffiltext{4}{\Goddard}
\altaffiltext{5}{\UBC}
\altaffiltext{6}{\PrincetonPhysics}
\altaffiltext{7}{\UChicago}
\altaffiltext{8}{\NRCFellow}
\altaffiltext{9}{\Brown}
\altaffiltext{10}{\UCLA}
\email{lverde@astro.princeton.edu}

\begin{abstract}

We describe our methodology for comparing the {\sl WMAP} measurements of the cosmic microwave background (CMB) and other complementary data sets to theoretical models. The unprecedented quality of the {\sl WMAP} data, and the tight constraints on cosmological parameters that are derived, require a rigorous analysis so that the approximations made in the modeling do not lead to significant biases.

We describe our use of the likelihood function to characterize the statistical properties of the microwave background sky.  We outline the use of the Monte Carlo Markov Chains to explore the likelihood of the data given a model to 
determine the best fit cosmological parameters and their uncertainties.

We add to the {\sl WMAP} data the  $\ell \ga 700$ CBI and ACBAR measurements of the CMB, the galaxy power spectrum at $z \sim 0$ obtained from the 2dF galaxy redshift survey (2dFGRS),  and the matter power spectrum at $z\sim 3$ as measured with the \lya forest. These last two data sets complement the CMB measurements by probing the matter power spectrum of the nearby universe.  Combining CMB and 2dFGRS requires that we include in our analysis a model for galaxy bias, redshift distortions, and the non-linear growth of structure. 
We show how the statistical and systematic uncertainties in the model and the data are propagated through the full analysis.
\end{abstract}

\section{INTRODUCTION}
CMB experiments are powerful cosmological probes because the early
universe is particularly simple and because the fluctuations
over angular scales $\theta>0\fdg2$ are described by linear theory
\citep{peebles/yu:1970,bond/efstathiou:1984,zaldarriaga/seljak:2000}. 
Exploiting this simplicity to obtain precise 
constraints on cosmological parameters requires that we accurately 
characterize the performance of the instrument 
\citep{jarosik/etal:2003b,page/etal:2003b,barnes/etal:2003,hinshaw/etal:2003b},
the properties of the foregrounds \citep{bennett/etal:2003c}, and 
the statistical properties of the microwave sky.

The primary goal of this paper is to present our approach to extracting the cosmological parameters from the temperature-temperature angular power spectrum (TT) and the  temperature-polarization angular cross-power spectrum (TE).
In companion papers, we present the TT \citep{hinshaw/etal:2003} and TE \citep{kogut/etal:2003}  angular power spectra and  show that the CMB fluctuations may be treated as Gaussian \citep{komatsu/etal:2003}.

Our basic approach is to constrain cosmological parameters with a likelihood
analysis first of the {\sl WMAP} TT and TE spectra alone, then  jointly with other CMB angular power spectrum determinations at higher angular resolution, and  finally of all CMB power spectra data jointly with the power spectrum of the large-scale structure (LSS). In \S \ref{sec:Like} we describe the use of the likelihood function for the analysis of microwave background data. This builds on the \citet{hinshaw/etal:2003} methodology for determining the TT spectrum and its curvature matrix, and  \citet{kogut/etal:2003} who describe our methodology for determining the TE  spectrum.  In \S \ref{sec:MCMC} we describe our use of Markov Chains Monte Carlo  (MCMC) techniques to evaluate the likelihood function of model parameters. While {\sl WMAP}'s measurements are a powerful probe of cosmology, we can significantly enhance their scientific value by combining the {\sl WMAP} data with other astronomical data sets.  This paper also presents our approach for including  external CMB data sets (\S \ref{sec:CMB}), LSS data (\S \ref{sec:LSS}) and \lya forest data (\S \ref{sec:Lya}). When including external data sets the reader should keep in mind that the physics and the instrumental effects  involved in the interpretation of these external data sets (especially 2dFGRS and \lya) are much more complicated and less well understood than for WMAP data. Nevertheless we aim to match the rigorous treatment of uncertainties in the {\sl WMAP} angular power 
spectrum with the inclusion of known statistical and systematic effects (of the data and of the theory), in the complementary data sets.

\section{LIKELIHOOD ANALYSIS OF {\sl WMAP} ANGULAR POWER SPECTRA}
\label{sec:Like}

The first goal of our analysis program is to determine the  values and confidence levels of the cosmological parameters that best describe the {\sl WMAP} data for a given cosmological model. We also wish to discriminate between different classes of cosmological models, in other words to assess whether a cosmological model is an acceptable fit to {\sl WMAP} data. 

The ultimate goal of the likelihood analysis is to find a set of parameters that give an estimate of $\langle {\cal C}_{\ell}\rangle$, the ensemble average of which  the realization on our sky\footnote{Throughout this paper we use the convention that
${\cal C}_{\ell}=\ell(\ell+1)C_{\ell}/(2\pi)$.} is ${\cal C}_{\ell}^{\rm sky}$. The likelihood function, ${\cal L}(\widehat{\cal C}_{\ell}|{\cal C}_{\ell}^{\rm th}(\vec{\alpha}))$, yields the probability of the data given a model and its parameters ($\vec{\alpha}$). In our notation $\widehat{\cal C}_{\ell}$ denotes our best estimator of ${\cal C}_{\ell}^{\rm sky}$ \citep{hinshaw/etal:2003} and ${\cal C}_{\ell}^{\rm th}$ is the theoretical prediction for angular power spectrum.
From Bayes' Theorem, we can split the expression for the probability of a model given the data as:
\begin{equation}
{\cal P}( {\bf \alpha}|\widehat{\cal C}_{\ell}) = 
\cal L(\widehat{\cal C}_{\ell}|{\cal C}^{\rm th}_{\ell}({\bf \alpha}))
{\cal P}(\bf{\alpha}),
\end{equation}
where ${\cal P}(\vec{\alpha})$ describes our {\it priors} on cosmological parameters and we have neglected a normalization factor that does not depend on the parameters . Once the choice of the priors are specified, our estimator of $\langle {\cal C}_{\ell}\rangle$ is given by ${\cal C}_{\ell}^{\rm th}$ evaluated at the maximum of ${\cal P}(\vec{\alpha}|\widehat{\cal C}_{\ell})$.

\subsection{Likelihood Function}
One of the generic predictions of inflationary models
is that fluctuations in the gravitational potential have 
Gaussian random phases. 
 Since the physics that governs the evolution of the temperature and metric fluctuations is linear, the temperature fluctuations are also Gaussian.  If we ignore the effects of non-linear physics at $z < 10$ and the effect of foregrounds, then all of the cosmological information in the microwave sky is encoded in the themperature and polarization power spectra. The leading-order low-redshift astrophysical effect is expected to be gravitational lensing of the CMB by foreground structures.  We ignore this effect here as it  generates a $<1$\% covariance in the TT angular power spectrum on {\sl WMAP} angular scales \citep{hu:2001b} (see Spergel et al. 2003, \S 3).
 
There are several expected sources of non-cosmological signal and of non-Gaussianity in the microwave sky.  The most significant sources on the full sky are Galactic foreground emission, radio sources, and galaxy clusters. \citet{bennett/etal:2003c} show that these contributions are greatly reduced {\it if} we restrict our analysis to a cut sky that masks bright sources and regions of bright Galactic emission. The residual contribution of these foregrounds is further reduced by the use external templates to subtract foreground emission from the Q, V and W band maps.  \citet{komatsu/etal:2003} find no evidence for deviations from Gaussianity on this template-cleaned cut sky.  While the sky cut greatly reduces foreground emission, it has the unfortunate effect of coupling multipole modes on the sky so that the power spectrum covariance matrix is no longer diagonal. 
 The goal of this section is to include this covariance in the likelihood function.

The likelihood function for the temperature fluctuations observed by a noiseless experiment with full  sky coverage has the form:
\begin{equation}
{\cal L}(\vec{T}|{\cal C}_{\ell}^{\rm th}) \propto \frac{{\rm exp} \left[-(\vec{T} {\bf S}^{-1} \vec{T})/2\right]}{\sqrt{\det {\bf S}}}\;,
\label{eq:n2}
\end{equation}
where $\vec T$ denotes our temperature map; and $S_{ij} = \sum_{\ell} (2 \ell+1) C_{\ell}^{\rm th} P_{\ell}(\hat n_i \cdot \hat n_j)/(4 \pi)$, where the $P_{\ell}$ are the Legendre polynomials and $\hat n_i$ is the pixel position on the map.  If we expand the temperature map in spherical harmonics: $T(\hat{n})=\sum_{\ell m}a_{\ell m}Y_{\ell m}$, then the likelihood function for each $a_{\ell m}$ has a simple form:
\begin{equation}
{\cal L}(\vec{T}| C_l^{\rm th})  \propto \prod_{\ell m} \frac{\exp
\left(-|a_{\ell m}|^2/(2C^{\rm th}_{\ell}\right))}{\sqrt{C^{\rm th}_{\ell}}}\;.
\end{equation}
Since we assume that the universe is isotropic, the likelihood function is independent of $m$. Thus,  we can sum over $m$ and rewrite the likelihood function as
\be  
-2\ln {\cal L}=\sum_{\ell}(2\ell+1)\left[\ln \left(\frac{{\cal
C}^{\rm th}_{\ell}}{\widehat{\cal C}_{\ell}}\right)+\widehat{\cal C}_{\ell}/{\cal C}^{\rm th}_{\ell}-1\right]
\label{eq:likelihoodexact}
\ee 
up to an irrelevant additive constant. Here, for a full sky, noiseless experiment  we have identified  $\sum_m|a_{\ell m}|^2/(2\ell+1)$ with $\widehat{C}_{\ell}$.
Note that the likelihood function depends only on the angular power spectrum. In this limit, the {\it angular power spectrum} encodes all of the cosmological information in the CMB.

Characteristics of the instrument are also included in the likelihood analysis.
\citet{jarosik/etal:2003} show that the detector noise is Gaussian (see their Figure 6 and \S 3.4); consequently the pixel noise in the sky map is also Gaussian \citep{hinshaw/etal:2003}.
The resolution of {\sl WMAP} is quantified with a window function, $w_{\ell}$ \citep{page/etal:2003}. Thus, the likelihood function for our CMB map has the same form as equation (\ref{eq:n2}), but with ${\bf S}$ replaced by
 ${\bf C} = \hat{\bf S} + {\bf N}$ where ${\bf N}$ is the nearly diagonal noise correlation matrix\footnote{$1/f$ noise makes a non-random phase contribution to the detector noise and leads to off-diagonal terms in the noise matrix. By making the noise $N_0$ a function of $\ell$ (denoted by $N_{\ell}$) we include this effect to leading order (Hinshaw et al. 2003)} and $\hat S_{ij}= \sum_{\ell} (2 \ell+1) C_{\ell}^{\rm th} w_{\ell} P_{\ell}(\hat n_i\cdot \hat n_j)/(4 \pi)$. 

If foreground removal did not require a sky cut and the noise were uniform and purely diagonal, then the likelihood function for the {\sl WMAP} experiment would have the form \citep{bond/jaffe/knox:2000}:
\begin{equation} 
-2\ln{\cal L}=\sum_{\ell}(2\ell+1)\left[\ln\left(\frac{{\cal C}_{\ell}^{\rm th}+{\cal N}_{\ell}}{\tilde{\cal C}_{\ell}}\right)+\frac{\tilde{\cal C}_{\ell}}{ {\cal C}_{\ell}^{\rm th} +{\cal N}_{\ell}}-1\right],
\label{eq:exact_like}
\end{equation}
\noindent where the effective bias ${\cal N}_{\ell}$  is related to the noise
bias $N_{\ell}$ as  ${\cal N}_{\ell}=N_{\ell}/w_{\ell}\ell(\ell+1)/(2\pi)$ and $\tilde{\cal C}_{\ell}=\ell(\ell+1)/(2\pi)\sum_m|a_{\ell m}|^2/(2\ell+1)/w_{\ell}$. 
Note that
${\cal N}_{\ell}$ and $\C_{\ell}^{\rm th}$ appear together in equation (\ref{eq:exact_like}) because the noise and cosmological fluctuations have the same statistical properties, they both are Gaussian random fields. 
  
Because of the foreground sky-cut, different multipoles are correlated and only a fraction of the sky, $f_{\rm sky}$, is used in the analysis. In this case, it becomes computationally prohibitive to compute the exact form of the likelihood function. There are several different approximations used in the CMB literature for the likelihood function.  At large $\ell$, equation (\ref{eq:exact_like}) is often approximated as Gaussian:
\be
\ln {\cal L}_{\rm Gauss}\propto -\frac{1}{2}\sum_{\ell\ell'}
({\cal C}_{\ell}^{\rm th}-\widehat{\cal C}_{\ell})Q_{\ell \ell'}
({\cal C}_{\ell'}^{\rm th}-\widehat{\cal C}_{\ell'})\;,
\label{eq:like_gauss}
\ee
where $Q_{\ell \ell'}$, the curvature matrix, is the inverse of the power spectrum covariance matrix.

The power spectrum covariance encodes the uncertainties in the power spectrum due to cosmic variance, detector noise, point sources, the sky cut, and systematic errors. \citet{hinshaw/etal:2003b} and \S (\ref{sec:curvature})  describe the various terms that enter into the power spectrum covariance matrix.

Since the likelihood function for the power spectrum is slightly non-Gaussian, equation (\ref{eq:like_gauss}) is a systematically biased estimator. \citet{bond/jaffe/knox:2000} suggest using a lognormal distribution, ${\cal L}_{\rm LN}$ \citep{bond/jaffe/knox:2000, sievers/etal:2002}:
\be 
-2\ln \cL_{\rm LN}=\sum_{\ell
\ell'}(z_{\ell}^{\rm th}-\widehat{z}_{\ell}){\cal Q}_{\ell
\ell'}(z_{\ell'}^{\rm th}-\widehat{z}_{\ell'}),
\label{eq:lognormallikelihood}
\ee
\noindent where $z_{\ell}^{\rm th}=\ln(\C_{\ell}^{\rm th}+{\cal N}_{\ell})$, 
$\widehat{z}_{\ell}=\ln(\widehat{\C}_{\ell}+{\cal N}_{\ell})$  and ${\cal Q}_{\ell \ell'}$ is the local transformation of the curvature matrix $Q$ to the lognormal variables $z_{\ell}$, 
\be 
{\cal Q}_{\ell
\ell'}=(\widehat{\cal C}_{\ell}+{\cal N}_{\ell})Q_{\ell \ell'}(\widehat{\cal C}_{\ell'}+{\cal N}_{\ell'}).
\label{eq:lognormalvariables}
\ee
We find that, for the {\sl WMAP} data, both equations (\ref{eq:like_gauss}) and (\ref{eq:lognormallikelihood}) are biased estimators. We use an alternative approximation of the likelihood function for the ${\cal C}_{\ell}$'s (equation \ref{eq:likelihoodform}) motivated by the following argument.

We can expand the exact expression for the likelihood (equation \ref{eq:likelihoodexact}) around its maximum by writing $\widehat{\cal C}_{\ell}={\cal C}^{\rm th}_{\ell}(1+\epsilon)$. Then, for a single multipole $\ell$,
\be 
-2\ln {\cal
L}_{\ell}=(2\ell+1)[\epsilon-\ln(1+\epsilon)]\simeq (2\ell+1)\left
(\frac{\epsilon^2}{2}-\frac{\epsilon^3}{3}+{\cal O}(\epsilon^4)\right)\;.
\ee 
We note that the Gaussian likelihood approximation is equivalent
to the above expression truncated at $\epsilon^2$: $-2 \ln {\cal L}_{{\rm Gauss},\ell}\propto(2\ell+1)/2 [(\widehat{\cal C}_{\ell}-{\cal C}_{\ell}^{\rm th})/{\cal C}_{\ell}^{\rm th}]^2\simeq (2\ell+1)\epsilon^2/2$. 

The \citet{bond/jaffe/knox:1998} expression for the lognormal likelihood for the equal variance approximation is
\be 
-2\ln {\cal L}'_{\rm LN}=\frac{(2
\ell+1)}{2}\left[\ln\left(\frac{\widehat{\cal C}_{\ell}} {{\cal
C}^{\rm th}_{\ell}}\right)\right]^2\simeq (2\ell+1)\left(\frac{\epsilon^2}{2}-
\frac{\epsilon^3}{2}\right)\;. 
\ee 
Thus our approximation of likelihood function is given by the form,
\be
\ln {\cal L}=\frac{1}{3} \ln {\cal L}_{\rm Gauss}+\frac{2}{3}\ln {\cal L}'_{\rm LN}\;,
\label{eq:likelihoodform}
\ee
where ${\cal L}'_{\rm LN}$ has the form of equation (\ref{eq:lognormallikelihood}) apart from ${\cal Q}_{\ell \ell'}$ that is not given by equation (\ref{eq:lognormalvariables}) but by
\be
{\cal Q}_{\ell \ell'}= ({\cal C}^{\rm th}_{\ell}+{\cal N}_{\ell})Q_{\ell \ell'}(\C_{\ell'}^{\rm th}+
{\cal N}_{\ell'}).
\label{eq:equalvariance}
\ee
We tested this form of the likelihood by making 100,000 full sky realizations of the TT angular power spectrum ${\cal C}_{\ell}^{\rm th}$.  For each realization, the maximum likelihood amplitude of fluctuations in the underlying model was found and the mean value was computed. Since we kept all other model parameters fixed, this one dimensional maximization was computationally trivial. The Gaussian approximation (equation \ref{eq:like_gauss}) was found to systematically overestimate the amplitude of the fluctuations by 
$\simeq 0.8$\%, while the lognormal approximation underestimates it by $\simeq 0.2$\%. Equation (\ref{eq:likelihoodform}) was found to be accurate to better than $0.1$\%.
%better than 0.1\%.

\subsection{Curvature Matrix}
\label{sec:curvature}
We obtain the curvature matrix in a form that can be used in the likelihood analysis from the power spectrum covariance matrix for $\widehat{\cal C}_{\ell}$ computed in  \citet{hinshaw/etal:2003b}.
The matrix is composed of several terms of the following form:
\be 
\Sigma_{\ell \ell'}=\sqrt{D_{\ell} D_{\ell'}} \left(\delta^K_{\ell \ell'}
- r_{\ell \ell'}\right)+\epsilon_{\ell \ell'},
\label{eq:fisheranalytic1}
\ee
where $\epsilon_{\ell \ell'}$ is the coupling introduced by the beam uncertainties and point sources subtraction ($\epsilon_{\ell \ell'}=0$ if $\ell=\ell'$), $\delta^{K}$ denotes the Kronecker delta function, and $D_{\ell}$ denotes the diagonal terms,
\be
D_{\ell}=2\frac{({\cal C}^{th}_{\ell} +{\cal N}_{\ell})^2}{(2\ell+1)f_{sky}^2}\:.
\label{eq:fisheranalytic2}
\ee 
The quantity $r_{\ell \ell'}$ encodes the mode coupling due to the sky-cut and is the dominant off-diagonal term (it is set to be $0$ if $\ell=\ell'$). The mode-coupling coefficient, $r_{\ell \ell'}$, is most easily defined in terms of the curvature matrix, $Q_{\ell \ell'}=D_{\ell}^{-1}\delta^K_{\ell \ell'}+r_{\ell \ell'}/\sqrt{D_{\ell}D_{\ell'}}$ (see Hinshaw et al. 2003\footnote{In this equation we have set to zero the beam and point sources uncertainties. This is because the coupling coefficient is computed for an ideal cut sky.}).

The sky cut has two significant effects on the power spectrum covariance matrix.  Because less data is used, the covariance matrix is increased by a factor of $f_{\rm sky}$. An additional factor of $f_{\rm sky}$ arises from the coupling to nearby $\ell$ modes. The additional term does not lead to a loss of information as nearby $\ell$ modes are slightly anti-correlated.
 
\citet{hinshaw/etal:2003} describe the beam uncertainty and point source terms included in ${\cal N}_{\ell}$ and $\epsilon_{\ell \ell'}$. The beam  and calibration uncertainties depend on the realization of the angular power spectrum on the sky ${\cal C}_{\ell}^{\rm sky}$, not on the theoretical angular power spectrum ${\cal C}_{\ell}^{\rm th}$, thus they should not change as, in exploring the likelihood surface, we change  ${\cal C}_{\ell}^{\rm th}$ in the expression for $D_{\ell}$. This differs from other approaches  (e.g., \cite{bridle/etal:2002}). Rescaling all the contributions to the off diagonal terms in the covariance matrix with ${\cal C}_{\ell}^{\rm th}$ is not correct and leads to a 2\% bias in our estimator of  $\langle{\cal C}_{\ell}\rangle$ which propagates, for example, into a $\sim 2$\% error on the matter density parameter $\Omega_m$ or $\sim 2$\% error on the spectral slope $n_s$.
 
We find the curvature matrix by inverting equation (\ref{eq:fisheranalytic1})
\be
Q_{\ell \ell'}=D^{-1}_{\ell}\delta^{K}_{\ell \ell'}-\frac{\epsilon_{\ell \ell'}}
{D_{\ell}D_{\ell'}}+\frac{r_{\ell \ell'}}{\sqrt{D_{\ell}D_{\ell'}}}\;,
\label{eq:curvature}
\ee
where we have assumed that the off-diagonal terms are small. 
For cosmological models that have ${\cal C}_{\ell}^{\rm th}$ very different from the best fit $\widehat{\cal C}_{\ell}$, equation \ref{eq:curvature} does not yield the inverse of (\ref{eq:fisheranalytic1}): in these cases the inversion of $\Sigma_{\ell \ell'}$ needs to be computed  explicitely.

We do not propagate the  {\sl WMAP} 0.5\% calibration uncertainty in the covariance matrix as this uncertainty  does not affect cosmological parameters determinations. This systematic  only affects the power spectrum amplitude constraint at the 0.5\% level, while the statistical error on this quantity is $\sim 10$\%.
 
\subsubsection{Calibration with Monte Carlo Simulations}\label{sec:fishercalibr} 

\begin{figure}
\epsscale{0.6}
\plotone{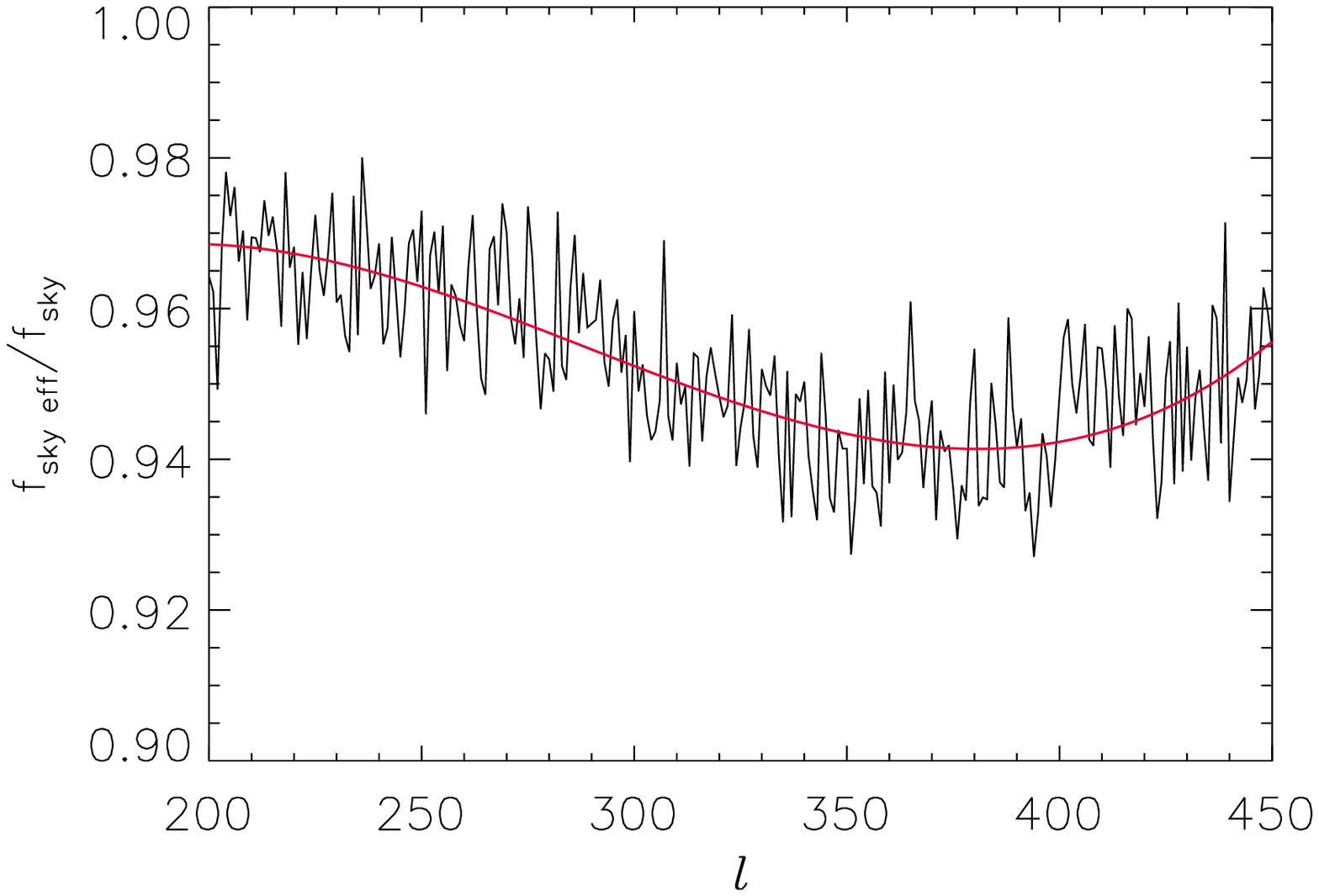}
\caption{Ratio of the effective sky coverage to the actual sky coverage. 
This correction factor calibrates the expression for the Fisher matrix to the value obtained from the Monte Carlo approach. Here we show the ratio obtained from 100,000 simulations (jagged line), the smooth curve shows the fit we use, equation (\ref{eq:fitfskyeff}). Note that, since we are switching between weighting schemes, the correction factors are not expected to smoothly interpolate between regimes.}
\label{fig.calibint}
\end{figure}

\begin{figure} 
\epsscale{0.6}
\plotone{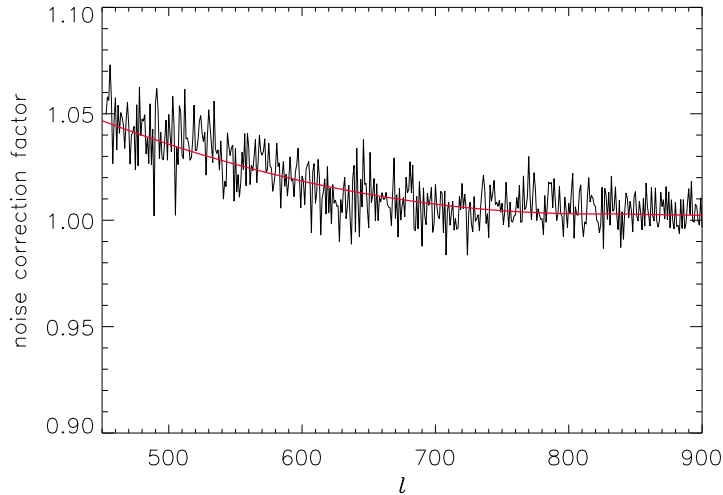}
\caption{Correction factor for the noise. The lines are as in Figure \ref{fig.calibint}.Note that, since we are switching between weighting schemes, the correction factors are not expected to smoothly interpolate between regime}
\label{fig.calibhighl}
\end{figure}

The angular power spectrum  is computed using three different weightings: uniform weighting in the signal-dominated regime ($\ell < 200$), an intermediate weighting scheme for $200 < \ell <450$, and $N_{obs}$ weighting (for the noise-dominated regime $450< \ell \le 900$ \citep{hinshaw/etal:2003}).  Uniform weighting is a minimum variance weighting in the signal-dominated regime  and $N_{obs}$ weighting is a minimum variance in the noise dominated regime. However, in the intermediate regime the weighting schemes are not necessarily optimal and the analytic expression for the covariance matrix might thus {\it underestimate} the errors.  To ensure that we have the appropriate errors, we calibrate the covariance matrix from 100,000 Monte Carlo realizations of the sky with the {\sl WMAP} noise level, symmetrized beams and the Kp2 sky cut. A good approximation of the curvature matrix can be obtained by using equations (\ref{eq:fisheranalytic1})--(\ref{eq:curvature}), but  substituting ${\cal N}_{\ell}$ and  $f_{\rm sky}$  with ${\cal N}_{\ell}^{\rm eff}$ and $f_{\rm sky}^{\rm eff}$  calibrated from the Monte Carlo simulations, as shown in Figures \ref{fig.calibint} and \ref{fig.calibhighl}.  

We find that for $\ell < 200$ the weighting scheme is nearly optimal. The power spectrum covariance matrix (\ref{eq:fisheranalytic1}) gives a correct estimate of the error bars, thus we do not need to calibrate ${\cal N}_{\ell}$ or $f_{\rm sky}$. We have computed an effective reduced chi-squared\footnote{This is not exactly the reduced chi-squared because the likelihood is non-gaussian especially at low $\ell$.} $\chi^2_{\rm eff}/\nu\equiv-2\ln {\cal L}/\nu$ where $\nu$ is the number of degrees of freedom. The effective reduced chi-squared from the Monte-Carlo simulations in this $\ell$ range is consistent with unity.

In the intermediate regime  our {\it ansatz}  power spectrum covariance matrix (equation (\ref{eq:fisheranalytic1})) slightly underestimates the errors. 
This can be corrected by computing the covariance matrix for an effective fraction of the sky $f_{\rm sky}^{\rm eff}$ as shown in Figure \ref{fig.calibint}. The jagged line is the ratio obtained from the Monte Carlo simulations,  the smooth curve shows the fit to $f_{\rm sky}^{\rm eff}$ we adopt,
\begin{equation}
\frac{f_{\rm sky}^{\rm eff}}{f_{\rm sky}}=0.813+0.001914\ell-7.405\times10^{-6}\ell^2+8.65\times10^{-9}\ell^3\;\;\;\; \mbox{(for $200<\ell<450$)}.
\label{eq:fitfskyeff}
\end{equation}

For $\ell > 450$, in the noise-dominated regime, the weighting is asymptotically optimal for $l \longrightarrow \infty$. However, since we are using a smaller fraction of the sky, we need again the correct the $f_{\rm sky}$ factor.  This numerical factor describes the reduction in effective sky coverage due to weighting the well observed ecliptic poles more heavily than the ecliptic plane (see Figure 3 of \cite{bennett/etal:2003}).  We fit this factor to the numerical simulations of the TT  spectrum covariance matrix. \citet{kogut/etal:2003} notes that this same factor is also a good fit to the Monte Carlo simulations of the TE spectrum covariance matrix.  For the noise-dominated regime, we define an effective sky fraction $f_{\rm sky}^{\rm eff}=f_{\rm sky}/1.14$ and an effective noise given by ${\cal N}^{\rm eff}_{\ell}=\sqrt{\Sigma^{\rm sim}_{\ell \ell} \left(f_{\rm sky}^{\rm eff}\right)^2(2 \ell+1)/2}-{\cal C}_{\ell}^{\rm sim}$, which can be obtained from the noise bias of the maps ${\cal N}_{\ell}$ by a noise correction factor ${\cal N}^{\rm eff}_{\ell}/{\cal N}_{\ell}$. This is shown in Figure \ref{fig.calibhighl} where the smooth curve is the fit we adopt to this correction factor, 
\begin{equation}
\frac{{\cal N}^{\rm eff}_{\ell}}{{\cal N}_{\ell}}=1.046-0.0002346(\ell-450)+3.204\times10^{-7}(\ell-450)^2\;\;\;\mbox{for $\ell >450$}\;. 
\end{equation}

This calibration of the covariance matrix from the Monte Carlo simulations allows us to use the effective reduced chi-squared as a tool to assess goodness of fit. It can also be used to determine the relative likelihood of different models (e.g., Peiris et al. 2003).

\subsection{Likelihood for the TE angular power spectrum}
Since the TE signal is noise-dominated, we adopt a Gaussian likelihood, where the curvature matrix is given by
\begin{equation}
Q^{TE}_{\ell \ell'}=(D^{TE}_{\ell})^{-1}\delta^K_{\ell \ell'}+r^{TE}_{\ell \ell'}/\sqrt{D^{TE}_{\ell}D^{TE}_{\ell'}}\;.
\label{eq:curvatureTE}
\end{equation}
The expression for $D^{TE}_{\ell}$ is given by  equation (10) of \citet{kogut/etal:2003}, and the coupling coefficient due to the sky cut, $r^{TE}_{\ell \ell'}$, is obtained from 100,000 Monte Carlo realizations of the sky with {\sl WMAP} mask and  noise level.  
The TE spectrum is computed with noise inverse weighting; in this regime 
$r_{\ell \ell'}$ depends only on the difference $\Delta_{\ell}=\ell-\ell'$ and is set to be 0 at separations $\Delta_{\ell}> 15$.
We use all multipoles $2\le \ell\le 450$, as  comparison with the Monte Carlo realizations shows  that in this regime equation (\ref{eq:curvatureTE}) correctly estimates the TE uncertainties. We have also verified  on the simulations that the Gaussian likelihood is an unbiased estimator, and that the effective reduced $\chi^2$ is centered around $1$. 

The amplitude of the covariance between TT and TE power spectra is $\sim \widetilde{r}/(1+n_{\ell}^{EE}/C_{\ell}^{EE})$ where $\widetilde{r}$ is the correlation term $(C_{\ell}^{TE})^2(C_{\ell}^{EE}C_{\ell}^{TT})^{-1}\simeq 0.2$. Since $C_{\ell}^{EE}/n_{\ell}^{EE}<<0.25$ for 1-yr data, we neglect this term, but we will include it in the 2+ yr analysis as it becomes increasingly important. 

We provide a subroutine that reads in a set of ${\cal C}^{\rm th}_{\ell}$ (TT, or TE or both) and returns the likelihood for the {\sl WMAP} dataset including all the effects described in this section. The routine is available at \verb"http://lambda.gsfc.nasa.gov".
 
\section{MARKOV CHAINS MONTE CARLO LIKELIHOOD ANALYSIS}
\label{sec:MCMC}

The analysis described in \citet{spergel/etal:2003} and 
\citet{peiris/etal:2003} is numerically demanding.
At each point in the six or more dimensional parameter space a new model 
from {\sf CMBFAST}\footnote{We used the parallelized 
version 4.1 of {\sf CMBFAST} developed in collaboration with 
Uros Seljak and Matias Zaldarriaga.} \citep{seljak/zaldarriaga:1996} 
is computed. Our version of the code 
incorporates a number of corrections and uses the {\sf RECFAST} 
\citep{seager/sasselov/scott:1999} recombination routine.   
Most of the likelihood calculations 
were done with four shared memory 32 CPU SGI Origin 300 with 
600 MHz processors.  With 8 processors per calculation, each 
evaluation of {\sf CMBFAST} for $\ell < 1500$ for a flat 
reionized $\Lambda$ dominated universe requires 3.6 seconds. (The
scaling is not linear; with 32 processors each evaluation requires 1.62
seconds.)

A grid-based likelihood analysis would have required prohibitive amounts of CPU
time. For example, a coarse grid ($\sim 20$ grid points 
per dimension) with six parameters requires 
$\sim 6.4\times 10^7$ evaluations of the power spectra. 
At 1.6 seconds per evaluation, the calculation would take $\sim 1200$
days. \cite{christensen/meyer:2000} proposed using Markov Chain Monte Carlo 
(MCMC) to investigate the likelihood space.  This approach has become the standard 
tool for CMB analyses \citep[e.g.,][]{christensen/etal:2001,knox/christensen/skordis:2001,
lewis/bridle:2002,kosowsky/milosavljevic/jimenez:2002}  and is the backbone of our 
analysis effort. 
For a flat reionized $\Lambda$ dominated universe, we can evaluate
the likelihood $\sim 120,000$ times in $< 2$ days 
using four sets of eight processors. 
As we explain below, this is adequate for finding the best fit model and for
reconstructing the 1- and 2-$\sigma$ confidence levels for the 
cosmological parameters.

We refer the reader to \citet{gilks:MCMCIP} for more information 
about MCMC. Here, we will only provide a brief introduction to 
the subject and concentrate on the issue of convergence.

\subsection{Markov Chain Monte Carlo}

MCMC is a method to simulate posterior distributions. In particular, we simulate observations from the posterior distribution ${\cal P}({\bf \alpha}|x)$, of a set of parameters ${\bf \alpha}$ given event $x$, obtained via Bayes' Theorem,
\begin{equation}
{\cal P}(\alpha|x)=\frac{{\cal P}(x|\alpha){\cal P}(\alpha)}{\int
{\cal P}(x|\alpha){\cal P}(\alpha)d\alpha},
\label{eq:bayes}
\end{equation}
\noindent where ${\cal P}(x|\alpha)$ is the likelihood of event $x$ given the model parameters $\alpha$ and ${\cal P}(\alpha)$ is the prior probability density.  For our application the {\sl WMAP} $\alpha$ denotes a set of cosmological parameters (e.g., for the standard, flat $\Lambda$CDM model these could be, the cold-dark matter density parameter $\Omega_c$, the baryon density parameter $\Omega_b$, the spectral slope $n_s$, the Hubble constant --in units of $100$ km s$^{-1}$ Mpc$^{-1}$)-- $h$, the optical depth $\tau$ and the power spectrum amplitude $A$), and event $x$ will be the set of observed $\widehat{\cal C}_{\ell}$.

The MCMC generates random draws (i.e. simulations) from the posterior distribution that are a ``fair'' sample of the likelihood surface. From this sample, we can estimate all of the quantities of interest about the posterior distribution (mean, variance, confidence levels). The MCMC method scales approximately linearly with the number of parameters, thus allowing us to perform likelihood analysis in a reasonable amount of time.

A properly derived and implemented MCMC draws from the joint posterior density ${\cal P}(\alpha|x)$ once it has converged to the stationary distribution.  The primary consideration in implementing MCMC is determining when the chain has  {\it converged}. After an initial {\it ``burn-in''} period, all further samples can be thought of as coming from the stationary distribution. In other words the chain has no dependence on the starting location. 
 
\begin{figure}
\begin{center}
\includegraphics[angle=90,scale=0.45]{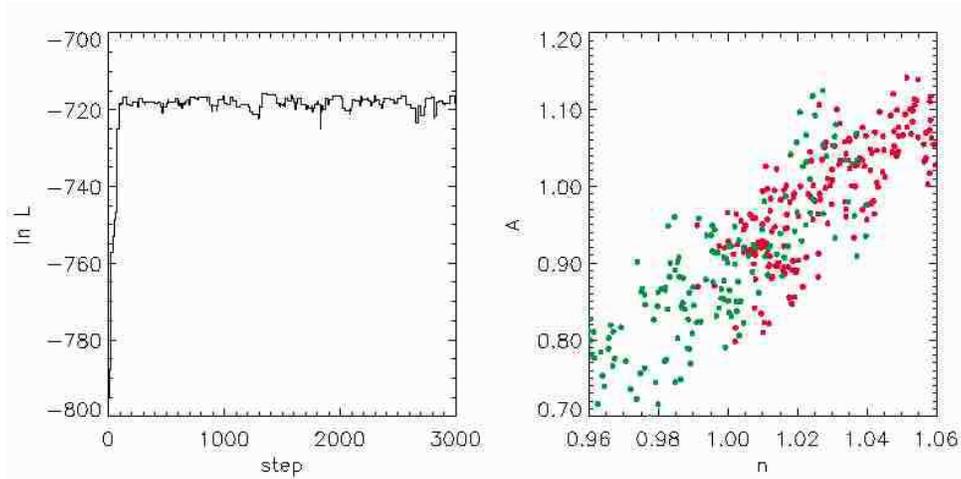}
\end{center}
\caption{Unconverged Markov chains. The left panel shown a {\it trace plot} of the likelihood values versus iteration number for one MCMC (these are the first 3000 steps  from one of our $\Lambda$CDM model runs). Note the {\it burn-in} for the first $\sim 100$ steps. In the right panel, red dots are points of the chain in the ($n$, $A$) plane after discarding the burn-in. Green dots are from another MCMC for the same data-set and the same model.   It is clear that, although the trace plot may appear to indicate that the chain has converged, it has not fully explored the likelihood surface. Using either of these two chains at this stage will give {\it incorrect} results for the best fit cosmological parameters and their errors.} 
\label{fig:unconv}
\end{figure}

Another fundamental problem of inference from Markov chains is that there are always  areas of the target distribution that have not been covered by a finite chain. If the MCMC is run for a very long time, the ergodicity of the Markov chain guarantees that eventually the chain will cover all the target distribution, but in the short term the simulations cannot tell us about areas where they have not been. It is thus crucial that the chain achieves good {\it ``mixing''}. If the
Markov chain does not move rapidly throughout the support of the target distribution because of poor {\it mixing}, it might take a prohibitive amount of time for the chain to fully explore the likelihood surface.  Thus it is important to have a convergence criterion and a mixing diagnostic.  Plots of the sampled MCMC parameters or likelihood values versus iteration number are commonly used to provide such criteria (left panel of Figure \ref{fig:unconv}). However, samples from a chain are typically serially correlated; very high auto-correlation leads to little movement of the chain and thus makes the chain to ``appear'' to have converged. For a more detailed discussion see \citet{gilks:MCMCIP}. Using a MCMC that has not fully explored the likelihood surface for determining cosmological parameters will yield {\it wrong} results.
We describe below the method we use to ensure convergence and good mixing.

\subsection{Convergence and Mixing}
\label{sec.conv}
We use the method proposed by \citet{gelman/rubin:1992} to
test for convergence and mixing, 
They advocate comparing several sequences drawn from different 
starting points and checking to see that they are indistinguishable. 
This method not only tests convergence but can also diagnose poor
mixing.
{\it For any analysis of the WMAP data,  we strongly encourage the use
of a convergence criterion.}

Let us consider $M$ chains (the analyses in \citet{spergel/etal:2003} and \citet{peiris/etal:2003} use 4 chains unless otherwise stated) starting at well-separated points in parameter space; each has $2N$ elements, of which we consider only the last N: $\{y_i^j\}$ where $i=1,..,N$ and $j=1,..,M$, i.e. $y$ denotes a chain element (a point in parameter space) the index $i$ runs over the elements in a chain the index $j$ runs over the different chains. We define the mean of the chain
\begin{equation}
\bar{y}^j=\frac{1}{N}\sum_{i=1}^{N}y_i^j,
\end{equation}
and the mean of the distribution
\begin{equation}
\bar{y}=\frac{1}{NM}\sum_{ij=1}^{NM}y_i^j\,.
\end{equation}
We then define the variance between chains as
\begin{equation}
B_n=\frac{1}{M-1}\sum_{j=1}^M(\bar{y}^j-\bar{y})^2,
\end{equation}
and the variance within a chain as
\begin{equation}
W=\frac{1}{M(N-1)}\sum_{ij}(y^j_i-\bar{y}^j)^2.
\end{equation}
The quantity
\begin{equation}
\hat{R}=\frac{\frac{N-1}{N}W+B_n\left(1+\frac{1}{M}\right)}{W}
\end{equation}
is the ratio of two estimates of the variance in the target distribution: the numerator is an estimate  of the variance that is unbiased if the distribution is stationary, but is otherwise an overestimate. The denominator is an underestimate of the variance of the target distribution if the individual sequences did not have time to converge.

The convergence of the Markov chain is then monitored by recording the quantity $\hat{R}$ for all the  parameters and running the simulations until the values for $\hat{R}$ are always $< 1.1$. Gelman (Kaas et al. 1997) suggest to use values for  $\hat{R} <1.2$.  Here, we conservatively adopt the criterion $\hat{R} < 1.1$ as our definition of convergence.
We have found that the four chains will sometimes go in and out of convergence as they explore the likelihood surface, especially if the number of points already in the chain is small. To avoid this, one could run many chains simultaneously or run one chain for a very long time (e.g., \cite{panter/heavens/jimenez:2002}). Due to CPU-time constraints,  we run four chains until they fulfill both of the following criteria  a) they have reached convergence, and b) each chain contains at least 30,000 points. In addition to minimizing chance deviations from convergence, we find that this many points are needed to be able to robustly reconstruct the 1-and 2-$\sigma$ levels of the marginalized likelihood for all the parameters.
For most chains, the burn-in time is relatively rapid, so that we only discard the first 200 points in each chain, however the results are not sensitive to this procedure.

\subsection{Markov Chains in Practice}

In this section we explain the necessary steps to run a MCMC for the CMB temperature power spectrum. It is straightforward to generalize these instructions to include the temperature-polarization power spectrum and other datasets. 
The MCMC is essentially  a random walk in parameter space, where the probability of being at any position in the space is proportional to the posterior probability.

Here is our basic approach:
\begin{itemize}
\item[1)] Start with a set of cosmological parameters $\{\alpha_1\}$, compute the ${\cal C}^{1}_{\ell}$ and the likelihood ${\cal L}_1={\cal L}({\cal C}^{1 {\rm th}}_{\ell}|\widehat{\cal C}_{\ell})$. 
\item[2)] Take a random step in parameter space to obtain a new set of cosmological parameters $\{\alpha_2\}$. The probability distribution of the step is taken to be Gaussian in each direction $i$ with r.m.s given by $\sigma_i$. We will refer below to $\sigma_i$ as the ``step size''. The choice of the step size is important to optimize the chain efficiency (see \S \ref{stepsizeopt})
\item[3)] Compute the ${\cal C}^{2 {\rm th}}_{\ell}$ for the new set of cosmological parameters and their likelihood ${\cal L}_2$.
\item[4.a)] If $\cL_2/\cL_1 \ge 1$, ``take the step'' i.e. save the new set of cosmological parameters $\{\alpha_2\}$ as part of the chain, then go to step 2 after the substitution $\{\alpha_1\}\longrightarrow \{\alpha_2\}$.
\item[4.b)] If $\cL_2/\cL_1 < 1$, draw a random number $x$ from a uniform distribution from 0 to 1. If $x \ge \cL_2/\cL_1 $ ``do not take the step'', i.e. save the parameter set $\{\alpha_1\}$ as part of the chain and return to step 2. If $x < \cL_2/\cL_1 $, `` take the step'', i.e. do as in 4.a).
\item[5)]For each cosmological model run four chains starting at randomly chosen, well-separated points in parameter space.  When the convergence criterion is satisfied and the chains have enough points to provide reasonable samples from the {\rm a posteriori} distributions (i.e. enough points to be able to reconstruct the 1- and 2-$\sigma$ levels of the marginalized likelihood for all the parameters) stop the chains. 
\end{itemize}
It is clear that the MCMC approach is easily generalized to compute the joint likelihood of {\sl WMAP} data with other datasets.

\subsection{Improving MCMC Efficiency}

The Markov chain efficiency can be improved in different ways.  We have tuned our algorithm by reparameterization and optimization of the step size.

\subsubsection{Reparameterization}

Degeneracies and poor parameter choices slow the rate of convergence and mixing of the Markov Chain. There is one near-exact degeneracy (the geometric degeneracy) and several approximate degeneracies in the parameters describing the CMB power spectrum \citep{bond/etal:1994, efstathiou/bond:1999}. 
The numerical effects of these degeneracies are reduced by finding a combination of cosmological parameters (e.g., $\Omega_c$, $\Omega_b$, $h$, etc.) that have essentially orthogonal effects on the angular power spectrum. The use of such parameter combinations removes or reduces degeneracies in the MCMC and hence speeds up convergence and improves mixing, because the chain does not have to spend time exploring degeneracy directions. \citet{kosowsky/milosavljevic/jimenez:2002} introduced a set of reparameterizations to do  just this. In addition, these new parameters  reflect the underlying physical effects determining the form of the CMB power spectrum (we will refer to these as physical parameters). This leads to particularly intuitive and transparent parameter dependencies of the CMB power spectrum.

Following \citet{kosowsky/milosavljevic/jimenez:2002}, we use a core
set of six physical parameters. There are two parameters for 
the physical energy densities of cold dark matter, $\omega_c\equiv \Omega_c
h^2$, and baryons, $\omega_b\equiv \Omega_b h^2$.
There is a parameter for the characteristic angular scale of the acoustic peaks,
\begin{equation}
\theta_A = \frac{r_s(a_{dec})}{d_A(a_{dec})},
\end{equation} 
\noindent where $a_{dec}$ is the scale factor at decoupling,
\begin{equation}
r_s(a_{dec})=\frac{c}{H_0\sqrt{3}}\int_0^{a_{dec}} 
\left[\left(1+\frac{3\Omega_b}{4\Omega_\gamma}\right)\left((1-\Omega)x^2+\Omega_\Lambda
x^{1-3w}+\Omega_m x + \Omega_{rad}\right)\right]^{-1/2}dx
\end{equation}
is the sound horizon at decoupling, and 
\begin{equation}
d_A(a_{dec})=\frac{c}{H_0}\int_{a_{dec}}^1
\left[(1-\Omega)x^2+\Omega_\Lambda
x^{1-3w}+\Omega_m x + \Omega_{rad}\right]^{-1/2}dx
\end{equation}
\noindent is the angular diameter distance at decoupling,  where $H_0$ denotes the Hubble constant and $c$ is the speed of light. Here
$\Omega_m=\Omega_c+\Omega_b$, $\Omega_{\Lambda}$ denotes  the vacuum energy density parameters, $w$ is the equation of state of the dark energy component,  $\Omega=\Omega_m+\Omega_{\Lambda}$ and  the radiation density parameter
$\Omega_{\rm rad}=\Omega_{\gamma}+\Omega_{\nu}$, $\Omega_{\gamma}$, $\Omega_{\nu}$ are the the photon and neutrino density parameters respectively. 
For reionization we use the physical parameter 
${\cal Z}\equiv \exp(-2\tau)$ where $\tau$ denotes the 
optical depth to the last scattering surface (not the decoupling
surface). 
The remaining two core
parameters are the spectral slope of the scalar primordial density 
perturbation power spectrum, $n_s$, and the overall amplitude of 
the primordial power spectrum $A$. Both are normalized at 
$k=0.05\,$ Mpc$^{-1}$ ($\ell \sim 700$). 

For more complex models we add other parameters as described in \citet{spergel/etal:2003} and \citet{peiris/etal:2003} and in \S \ref{sec:LSS}.
To investigate non-flat models
we use the vacuum energy, $\omega_{\Lambda}\equiv \Omega_{\Lambda}h^2$.
Other examples include
the tensor index, $n_t$, the tensor to scalar ratio, $r$, and the running of the scalar spectral index, $dn_s/d\ln k$.  

Here, we relate the input parameter for the overall normalization,
$A$, as in the {\sf CMBFAST} code (version 4.1 with UNNORM option),
to the amplitude of primordial comoving curvature perturbations ${\cal R}$,
$\Delta^2_{\cal R}(k_0)\equiv (k^3/2\pi^2)\langle|{\cal R}|^2\rangle$. 
We also relate our convention for the tensor perturbations to the one
in the code. {\sf CMBFAST} calculates
\begin{eqnarray}
  \label{eq:ClPsi}
  C^S_l &=& (4\pi) T_0^2 \int \frac{dk}k \Delta^2_{\Psi}(k) 
  \left[g_{Tl}^\Psi(k)\right]^2,\\
  \label{eq:ClT}
  C^T_l &=& (4\pi) T_0^2 \int \frac{dk}k \tilde{\Delta}^2_h(k)
  \left[g_{Tl}^h(k)\right]^2,
\end{eqnarray}
where $\Psi$ is the Newtonian potential,  
$g_{Tl}(k)$ is the radiation transfer function,
and $T_0=2.725\times 10^6$ is the CMB temperature in units of $\mu{\rm K}$.
The tilde denotes that $\tilde{\Delta}^2_h(k)$ is used in {\sf CMBFAST}, 
but differs from our convention, $\Delta_h^2(k)$, where $\tilde{\Delta}_h^2=\Delta_h^2/16$.
The comoving curvature perturbation, ${\cal R}$, is related to 
$\Psi$ by $\Psi=-(3/5){\cal R}$; thus, 
$\Delta^2_{\cal R}(k)= (25/9)\Delta^2_{\Psi}(k)$.
Note that this relation holds from radiation domination to matter
domination with accuracy better than 0.5\%.

{\sf CMBFAST} uses $A$ to parameterize $\Delta_\Psi^2(k_0)$.
The tensor perturbations are calculated accordingly.
The relations are
\begin{eqnarray}
   \Delta^2_{\Psi}(k_0) &=& \frac{800\pi^2}{T^2_0} A,\\
   \tilde{\Delta}^2_h(k_0) &=& 
    \frac1{16}\Delta^2_h(k_0)
    =\frac{r}{16} \Delta^2_{\cal R}(k_0)
   = \frac{25r}{144}  \Delta^2_{\Psi}(k_0).
\end{eqnarray}
Therefore, one obtains
\begin{equation}
 \label{eq:results}
  \Delta_{\cal R}(k_0)
  = 2.95\times 10^{-9} A.
\end{equation}

The  amplitude $A$ is normalized at $k_1=0.05$ Mpc$^{-1}$ and the tensor 
to scalar ratio $r$ is  evaluated at  $k_0=0.002$ Mpc$^{-1}$, 
unless otherwise specified.
To convert $A(k_0)$ to $A(k_1)$, we use
\be
\label{eq:conv_A}
  A(k_1)=A(k_0)
  \left(\frac{k_1}{k_0}\right)^{n_s(k_0)-1+\frac{1}{2}(dn_s/d\ln
  k)\ln(k_1/k_0)}.  
\ee

\subsubsection{Step Size Optimization}
\label{stepsizeopt}

The choice of the step size in the Markov Chain is crucial to improve the chain efficiency and speed up convergence. If the step size is too big, the acceptance rate will be very small; if the step size is too small the acceptance rate will be high but the chain will exhibit poor mixing. Both situations will lead to slow convergence.  For our initial step sizes for each parameter we use the standard deviation for each parameter when all the other parameters are held fixed at the maximum likelihood value. These are easy to find once a preliminary chain has been run and the likelihood surface has been fitted, as explained in \S\ref{likelihoodfitting}. If a given parameter is roughly orthogonal to all the other parameters, it is not necessary to adjust the step size further; in the presence of severe degeneracies the step size estimate needs to be increased by a ``banana correction'' factor which is approximately the ratio of the projection of the 1-$\sigma$ error along the degeneracy to the projection perpendicular to the degeneracy.

With these optimizations the convergence criterion is satisfied for the 4 chains after roughly  30,000 steps each ($2N=30,000$) for a model with 6 parameters. On a Origin 300 machine this takes roughly 32 hrs running each chain on  8 processors. These numbers serve only as a rough indication: convergence  speed depends on the model and on the data-set: for a fixed  number of parameters, convergence can be significantly slower if there are severe degeneracies among the parameters; adding more datasets might slow down the evaluation of a single step in the chain, but can  also speed up convergence by breaking degeneracies.

\subsubsection{Likelihood Surface Fitting}
\label{likelihoodfitting}

The likelihood surface explored by the MCMC was found to be 
functionally well approximated by a quartic expansion of the 
cosmological parameters (for example, 
$\{\alpha_i\} = \{\omega_b,\omega_c,n,\theta_A,{\cal Z},A\}$):
\begin{equation}
y \equiv \log(\cL) = q_0 + \sum_{i}{q_1^i \delta_i} + \sum_{i\le
j}{q_2^{ij} \delta_i \delta_j} + \sum_{i\le j\le k}{q_3^{ijk} \delta_i
\delta_j \delta_k} + \sum_{i\le j\le k\le l}{q_4^{ijkl} \delta_i
\delta_j \delta_k \delta_l} .
\label{eqn:lhfit}
\end{equation}
Here  $q$ are fit coefficients and  $\delta_i$ are related 
to the cosmological parameters via 
$\delta_i = (\alpha_i-\alpha^0_i)/\alpha_i$, 
where $\alpha^0_i$ is the maximum-likelihood value of 
the parameter.  Lower-order expansions were unable to reproduce 
the likelihood surface.  With 6 parameters there are $M_f=210$ fit 
coefficients.  Writing (\ref{eqn:lhfit}) as 
$y = {\vec q}\cdot{\vec x}$, the minimum least-squares 
estimator for $\vec q$ is
\begin{equation}
{\vec q} = (X^TX)^{-1}X^T {\vec y},
\end{equation}
where $X$ is the $N\times M_f$ matrix $X_{ij} = x^{(i)}_j$, $N$ the 
number of unique points in the chain.

We run preliminary MCMC chains with ``guesstimated'' step sizes 
until there are $\sim 1000$ unique points in total. 
Then we use equation~\ref{eqn:lhfit} to cut through the likelihood 
surface at the maximum likelihood value
to find the  $1\sigma$ level in each parameter 
direction (see \S~\ref{stepsizeopt}). This defines our ``step size''
for subsequent chains.

\subsection{The Choice of Priors}

From Bayes' Theorem (equation \ref{eq:bayes}) we can infer ${\cal P}(\alpha_i|x)$, the probability of the model parameters $\alpha_i$  given the event $x$ (i.e. our observation of the power spectra), from the likelihood function once the prior is specified. It is reasonable to take prior probabilities to be equal when nothing is known to the contrary (Bayes' postulate). Unless otherwise stated we assume uniform priors on the parameters given in Table 1.
Note that we assume uniform priors on $\omega_c$, $\omega_b$, and $\theta_A$ rather than uniform priors on $\Omega_m$, $\Omega_b$ and $H_0$. 

\begin{deluxetable}{c}
\tablecaption{Priors for Bayesian Analyses\label{tab:priors}}
\tablewidth{0pt}
\tablehead{
\colhead{Item}}
\startdata
$0\le \omega_c\le 1$\\
$0\le \omega_b\le 1$\\
$0.005\le {\theta_A}\le 0.1$\\ 
$0 \le \tau \le 0.3$\\ 
$0.5 \le A \le 2.5$\\
$0 \le n_s|_{k_0}\le 2$\\
$0 \le n_{\rm iso}\le 2$\\
$0 \le f_{\rm iso}\le 5000$\\
$-0.5 \le d n/d\ln k\le 0.5$\\ 
$0 \le r\le 2.5$\\
$0\le \omega_{\nu}\le 1$\\
$-3.2 (-1.2) $\tablenotemark{a} $\le  w \le 0$\\
$0\le \omega_{\Lambda}\le 1$\\ 
\enddata
\tablenotetext{a}{We will present two sets of results, one with  the prior $w\ge -1.2$ the other with $w\ge-3.2$}
\end{deluxetable}

Except for the priors on $\tau$, and $w$ (the equation of state of the dark energy component), the MCMCs never hit the imposed boundaries, thus most of our choices for priors have no effect on the outcome. A detailed discussion about the prior on $\tau$ is presented in Spergel et al. (2003). 
We set  lower bound on $w$ at $-3.2$ ($-1.2$) but we  discard the region of parameter space where $w<-3$ ($w<-1$). This is necessary because our best-fit value for this parameter is close to the boundary. 
If we had instead set the prior to
be $w\geq-3$ ($w\geq-1$), then the chains would fail to be a fair representation 
of the posterior distribution in the region of parameter space 
where the distance from the boundary is comparable to the stepsize.

\subsection{MCMC Output Analysis}
We merge the 4 converged MCMC chains ($\ga$  120,000 points) into one.
From this we give  the cosmological parameters that yield our  best estimate of ${\cal C}_{\ell}$ and  we give the marginalized distribution of the parameters.
We compute the marginalized distribution for one parameter, and the joint distribution for two parameters, obtained marginalizing over all the other parameters. Since the MCMC passes objective tests for convergence and mixing, the density of points in parameter space is proportional to the posterior probability of the parameters. 

The marginalized distribution is obtained by projecting the MCMC points.
For the marginalized parameters values $\bar{\alpha_i}$, \citet{spergel/etal:2003} quote the expectation value of the marginalized likelihood, $\int{\cal L}\alpha_id\alpha_i=1/N\sum_{t}\alpha_{t,i}$. Here, $N$ is the number of points in the merged chain and $\alpha_{t,i}$ denotes the value of parameter $\alpha_i$ at the $t$-th step of the chain.  The last equality becomes clear if we consider that the MCMC gives to each point in parameter space a ``weight'' proportional to the number of steps the chain has spent at that particular location. The 100(1-2p)\% confidence interval $[c_p,c_{1-p}]$ for a parameter is estimated by setting  $c_p$ to the $p^{th}$ quantile of $\alpha_{t,i},t=1,\ldots,N$ and $c_{p-1}$ to the $(1-p)^{th}$ quantile. 
The procedure is similar for multidimensional constraints: the density of points in the n-dimensional space is proportional to the likelihood and  multi-dimensional confidence levels can be found as illustrated in \S 15.6 of  Press et al. (1992). 
 
We note that the global maximum likelihood value for the parameters does not necessarily coincide with the expectation value of their marginalized distribution if the likelihood surface is not a multi-variate Gaussian.
We find that, for most of the parameters, the maximum likelihood values of the global joint fit are consistent with the expectation values of the marginalized distribution.

A virtue  of the MCMC method is that the addition of extra data sets in the joint analysis can efficiently be done with minimal computational effort from the MCMC output if the inclusion of extra data set does not require the introduction of extra parameters or  does not drive the parameters significantly away from the current best fit. For example, we add  \lya power spectrum constraint to  MCMC's outputs, but we cannot do this for the 2dFGRS, since this requires the introduction of two extra parameters ($\beta$, $\sigma_p$, see \S \ref{sec:2dF} below for more details).

If the likelihood surface for a subset of parameters from an external (independent) data set is known, or if a prior needs to be added {\it a posteriori}, the joint likelihood surface can be obtained by multiplying the likelihood with the posterior distribution of the MCMC output. In \citet{spergel/etal:2003}  we follow this method to obtain the joint constraint of CMB with Supernovae Ia  \citep{riess/etal:1998,riess/etal:2001} data  and CMB with Hubble Key project Hubble constant \citep{freedman/etal:2001}  determination.  

There is yet another advantage of the MCMC technique. 
The current version of {\sf CMBFAST} with the nominal interpolation settings is
accurate to 1\%, but {\it random} numerical errors can sometimes exceed this. As the precision of the CMB measurements improve, these effects can become problematic for any approach that calculates derivatives as a function of parameters.  Because
MCMC calculations average over $\sim$100,000 CMB calculations, the MCMC technique is much less sensitive than either grid-based likelihood calculations
or methods that numerically calculate the Fisher matrix.

\section{EXTERNAL CMB DATA SETS}
\label{sec:CMB}

The CBI \citep{mason/etal:2002, sievers/etal:2002, pearson/etal:2002} and the
ACBAR \citep{kuo/etal:2002} experiments complement {\sl WMAP} by probing the amplitude of CMB temperature power spectrum at $\ell > 900$.
These observations probe the Silk damping tail and improve our analysis in 2 ways: a) improve our ability to constrain the baryon density, the amplitude of fluctuations and the slope of the matter power spectrum, and b) improve convergence by preventing the chains from spending long periods of time in large, moderately low-likelihood regions of parameter space.

The CBI data set is described in Mason et al. (2002), Pearson et al. (2002) and on their web site\footnote{http://www.astro.caltech.edu/$\sim$tjp/CBI/data/index.html (last update August 2002)}. We use data from the CBI {\it mosaic} data set \citep{pearson/etal:2002} and do not include the deep data set as the two data sets are not independent. We use the three bandpowers from the even binning at central $\ell$ values of 876, 1126, 1301, thus  ensuring that the chosen bands power can be considered independent from the {\sl WMAP} data. At $\ell \ga 1500$, the CBI experiment detected excess power.  If the rms amplitude of mass fluctuations on scales of $8$ h$^{-1}$ Mpc, is $\sigma_8 \sim 1$, then this excess power can be interpreted as due to Sunayev-Zeldovich distortion from undetected galaxy clusters \citep{mason/etal:2002,bond/etal:2002,komatsu/seljak:2002}.  We simplify our analyses by not using the CBI data on scales where this effect can be important. The correlations between different band powers are taken into account with the full covariance matrix; we use the lognormal form of the likelihood (as in Pearson et al 2002). In addition, we marginalize over a 10\% calibration  uncertainty (CBI beam uncertainties are negligible).

\begin{figure}
\epsscale{0.6}
\plotone{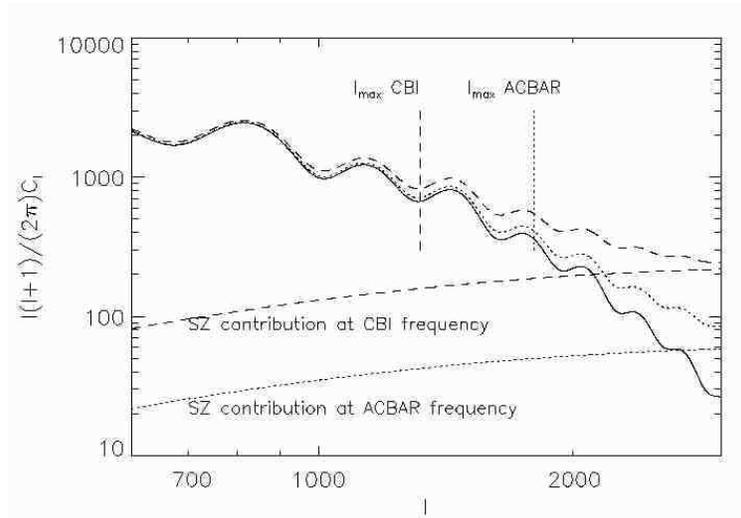}
\caption{The CMB angular power spectrum (in $\mu$K$^2$) for our best fit $\Lambda$CDM model for $\ell >800$ and the Sunayev-Zel'dovich contribution for $\sigma_8=0.98$ for CBI wavelengths (dotted) and for ACBAR (dashed). The vertical line shows the adopted cutoff for CBI and ACBAR.}
\label{fig:CMBext}
\end{figure}

The ACBAR data set is described in \citet{kuo/etal:2002}. We use the 7 band-powers at  multipoles 842, 986, 1128, 1279, 1426, 1580, 1716. As shown in Figure \ref{fig:CMBext}, these points do not overlap with the {\sl WMAP} power spectrum except at $\ell \sim 800$ where {\sl WMAP} is noise-dominated. As shown in Figure \ref{fig:CMBext} the ACBAR experiment is less sensitive to Sunyaev-Zel'dovich contamination than CBI. We compute the likelihood analysis for cosmological parameters for the ACBAR data set following \citet{goldstein/etal:2002} and using the error bars given in ACBAR web site\footnote{http://cosmology.berkeley.edu/group/swlh/acbar/data/}. In addition we marginalize over conservative beam and calibration uncertainties (B. Holzapfel 2002, private communication). In particular we assume a calibration uncertainty of 20\% (the double of the nominal value)  and 5\% beam uncertainty (60\% larger than the nominal value).  

The ACBAR and CBI data are completely independent from each other (they map different regions of the sky) and from the {\sl WMAP} data (the band-powers we consider span different $\ell$ ranges). To perform the joint likelihood analysis, we simply multiply the individual likelihoods.

\section{ANALYSIS OF LARGE SCALE STRUCTURE DATA}
\label{sec:LSS}

We can enhance the scientific value of the CMB data from $z \sim 1089$  by combining it with measurements of the low redshift universe.  Galaxy redshift surveys allow us to measure the galaxy power spectrum at $z \sim 0$ and observations of \lya absorption of about 50 quasar spectra (\lya forest) allow us to probe the dark matter power spectrum at redshift $z \sim 3$.

\begin{figure}
\epsscale{0.6}
\plotone{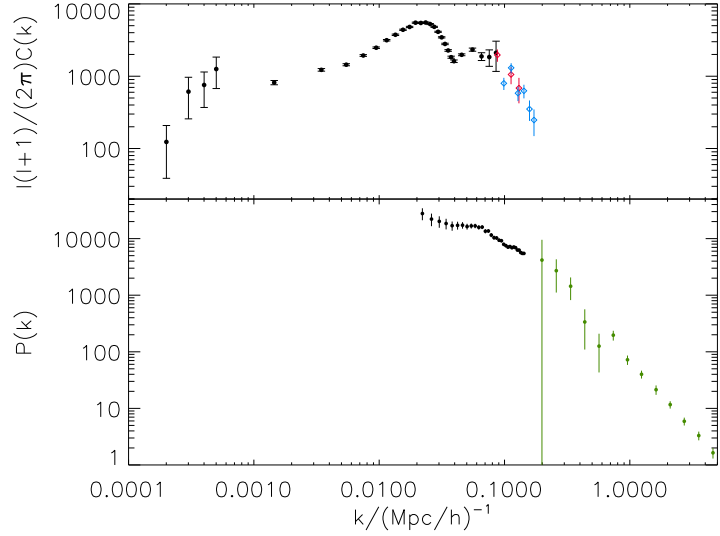}
\caption{The combined CMB and LSS data set. The CMB angular power spectrum in $\mu$K$^2$ (top panel) as a function of $k$ where $k$ is related to $\ell$ by $\ell=\eta_0 k$ (where $\eta_0 \sim 14400$ Mpc is the distance to the last scattering surface). Black points are the {\sl WMAP} data, red points CBI, blue points ACBAR. The LSS data (bottom panel). Black points are the 2dFGRS measurements and green points are the \lya measurements. Both LSS power spectra are in units of (Mpc $h^{-1}$)$^3$ and  have been rescaled to $z=0$. This plot only illustrates the scale coverage of all the data sets we consider. The various LSS power spectra as plotted here cannot be directly compared with the theory because of  the effects outlined in \S 5 (e.g., redshift-space distortions, non-linearities, bias and window function effect etc.).} 
\label{fig:alldata}
\end{figure}

We use the Anglo-Australian Telescope Two Degree Field Galaxy Redshift
Survey (2dFGRS) \citep{colless/etal:2001} as compiled in February 2001. This survey probes the universe at redshift $z_{\rm eff}\sim 0.1$ and probes the power spectrum on scales corresponding to $0.022<k<0.2$ (where  $k$ is in units of h Mpc$^{-1}$.  The anticipated Sloan Digital Sky Survey  \citep{gunn/knapp:1993} power spectrum will be an important complement to 2dFGRS.  We also use the linear matter power spectrum as recovered by \citet{croft/etal:2002} from \lya forest observations. This power spectrum is reconstructed at an effective redshift $z \sim 2.72$ and probes scales $k>0.2$ h Mpc$^{-1}$.
Together these data sets allow us not only to probe a wide range of physical scales---from $k\sim 1\times 10^{-4}$ ($30000$ Mpc h$^{-1}$) to $k \sim 1$ ($3$ Mpc h$^{-1}$)--- (see Figure \ref{fig:alldata}),  but also  to probe the evolution of a given scale with redshift as well.

When including LSS data sets one should keep in mind that the underlying physics for these data sets is much more complicated and less well understood than for {\sl WMAP} data, and systematic and instrumental effects are much more important. We  attempt here to include all the known (up to date) uncertainties and systematics in our analysis.
In what follows, we illustrate our modeling of the ``real-world'' effects of LSS surveys and how we propagate systematic and statistical uncertainties into the parameters estimation. The goal of our modeling is to relate not just the {\it shape} but also the {\it amplitude} of the observed power spectrum to that of the linear matter power spectrum as constrained by CMB data. The reason for this will be clear in \S \ref{sec:motivation}; by using the information in the power spectrum amplitude we can break some of the degeneracies among cosmological parameters. 

\subsection{The 2dFGRS Power Spectrum}
\label{sec:2dF}

The 2dFGRS power spectrum, as released in June 2002, has been calculated from the February 2001 catalog that includes 140,000 galaxies \citep{percival/etal:2001}. 
The full survey is composed of 220,000 galaxies
but is not yet available.  The sample is magnitude-limited at $b_{J}=19.45$ and thus probes the universe at $z_{\rm eff}\sim 0.1$ and the power spectrum on scales corresponding to $k>0.015$ h Mpc$^{-1}$.  The input catalog is an extended version of the Automatic Plate Machine (APM) galaxy catalog \citep{maddox/etal:1990,maddox/efstathiou/sutherland:1990, maddox/efstathiou/sutherland:1996} which includes about 5 million galaxies  to $b_{J}=20.5$. The APM catalog was used previously to recover the 3-d power spectrum of galaxies by inverting the clustering properties of the 2-d galaxy distribution \citep{baugh/efstathiou:1993, efstathiou/moody:2001}. These techniques, however, are affected by sample variance and uncertainties in the photometry; a full 3-d analysis is thus more reliable.

The power spectrum of the galaxy distribution as measured by LSS surveys, such as the 2dFGRS, cannot be directly compared to that of the initial density fluctuations as predicted by theory, or recovered from {\sl WMAP} or, the combination of  {\sl WMAP}+CBI+ACBAR data-sets. This is due to a number of intervening effects that can be broadly divided in two classes: effects due to the survey geometry (i.e., window function, selection function effects) and effects intrinsic to the galaxy distribution (e.g., redshift-space distortions, bias, non-linearities).

\subsubsection{Survey Geometry}
\label{sec:surveygeom}

Galaxy surveys such as the 2dFGRS are magnitude-limited rather than volume-limited, thus most nearby galaxies are included in the catalog while only the brighter of the more distant galaxies are selected. The selection function accounts for the fact that fewer galaxies are included in the survey as the distance (or the redshift) increases. An additional effect arises from the fact that the clustering properties of bright galaxies might be different from the average clustering properties of the galaxy population as a whole. The selection function does not take this into account (we will return to this point in \S \ref{sec:galdistr}).

Moreover, the completeness across the sky is not constant and the survey can only cover a fraction of the whole sky, sometimes with a very complicated geometry described by the window function. In particular, for the data we use, unobserved fields make the survey completeness a strongly varying function of position.
The measured Fourier coefficients are therefore the true coefficients of the galaxy distribution convolved by the Fourier transform of the selection function (in the direction of the line of sight) and of the window function (on the plane of the sky). In this section, we follow the standard notation used in LSS analyses and refer to all of these effects as window effects.

The window not only modifies the measured power spectrum but also introduces spurious correlations between Fourier modes. (See Percival et al. (2001) for more details). For the 2dFGRS these effects have been quantified by Monte Carlo simulations of mock catalogs of the survey\footnote{For {\sl WMAP} data, we deconvolve the raw measured $\tilde{C}_{\ell}$ by the effect of the window (the mask), thus leaving the effect of the window function and the mask only in the fisher matrix. For LSS we will convolve the theory with the window, project the power spectrum into redshift space and compare this to the observed power spectrum.}.
We include them in our analysis by convolving the theory power spectrum with the window ``kernel'', and by including off diagonal terms in the covariance matrix. 

\subsubsection{Effects Intrinsic to the Galaxy Distribution}
\label{sec:galdistr}

\begin{figure}
\epsscale{0.5}
\plotone{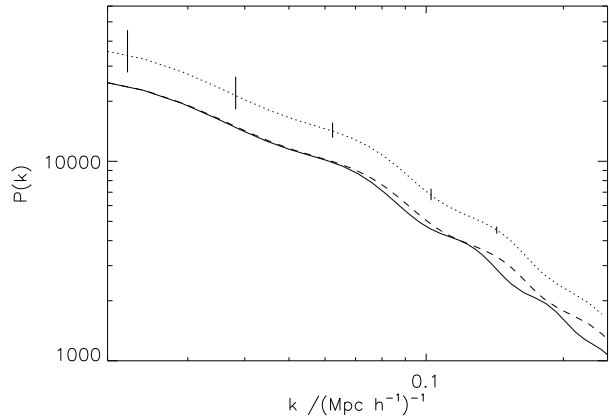}
\caption{The matter power spectrum in (Mpc $h^{-1}$)$^3$), linear in real space (solid line), non-linear in real space (dashed line) and non-linear in redshift space (dotted line). The error bars on the dotted line show the size of the statistical error-bars per $k$-bin of the 2dFGRS galaxy power spectrum. The power spectrum is in units of (Mpc $h$)$^{3}$.} 
\label{fig:powerspredsh}
\end{figure}

Linear gravitational evolution modifies the amplitude but not the shape of the underlying power spectrum.
However, in the non-linear regime (where the amplitude of fluctuations is $\delta\rho/\rho \sim 1$) this is no longer the case. Non-linear gravitational evolution changes the shape of the power spectrum and introduces correlations between Fourier modes. This effect becomes important on scales $k \sim 0.1$ h Mpc$^{-1}$,  but the exact scale at which it appears and its detailed characteristic depend on cosmological parameters. Most of the clustering signal from galaxy surveys such as 2dFGRS comes from the regime where non-linearities are non-negligible because shot noise is the dominant source of error  at $k \ga 0.5$ h Mpc$^{-1}$  and the number density of modes scales as $k^3$. These non linearities encode additional information about cosmology and motivates their inclusion in the present analysis. This approach is complicated by the fact that an accurate description of the fully non-linear evolution of the {\em galaxy} power spectrum is complicated. In the literature, there are several different approaches to modeling the non-linear evolution of the underlying {\em dark matter} power spectrum in real space: (1) linear (and extended) perturbation theory; (2) semi-analytical modeling and (3) numerical simulation.  All of these approaches yield consistent results on the scales used in our analysis. We will use the semi-analytical approach developed by \citet{hamilton/etal:1991} and \citet{peacock/dodds:1996}.   In particular, we use the \citet{ma/etal:1999} formulation of the  non-linear power spectrum. Figure \ref{fig:powerspredsh} shows the effect of non-linearities on the matter power spectrum on the scales of interest (compare solid and dashed lines).

Theory predicts the statistical properties of the continuous matter distribution, while observations are concerned with the galaxy distribution, which is discrete. Moreover, galaxies might not be faithful tracers of the mass distribution (i.e. the galaxy distribution might be {\it biased}). In the analysis of galaxy surveys it is assumed that galaxies form a {\it Poisson sampling} of an underlying continuous field which is related to the matter fluctuation field via the {\it bias}. It is possible to formally relate the discrete galaxy field and its continuous counterpart. For the power spectrum, this consists  of the subtraction from the measured galaxy power spectrum of the shot noise contribution. The published power spectra from galaxy surveys already have this contribution subtracted, but are still biased with respect to the underlying mass power spectra.

The idea that galaxies are biased tracers of the mass distribution even on large scales was introduced by \citet{kaiser:1984} to explain the properties of Abell clusters. Nevertheless, the fact that galaxies of different morphologies have different clustering properties (hence different power spectra) was known much before (e.g., \citet{hubble:1936,dressler:1980,postman/geller:1984}). Since the clustering properties of different types of galaxies are different, they cannot all be good tracers of the underlying mass distribution\footnote{Galaxies are likely to be formed in the very high-density regions of the matter fluctuation field, thus they are formed very biased at $z >> 0$ (e.g., Lyman break galaxies). But then gravitational evolution should make the galaxy distribution less and less biased as time goes on (e.g., \citet{fry:1996})} .  

In the simplest biasing model, the linear bias model, the mass and galaxy fractional overdensity fields $\delta$ and $\delta_g$ are related by
$ \delta_g({\bf x})=b\delta({\bf x})$. This implies that on all scales 
\be
P_g(k)=b^2P(k).
\label{eq:biasedpowersp}
\ee
This simple model (although justified by the Kaiser (1984) assumption that galaxies form on the highest peaks of the mass distribution) cannot be true in detail for two reasons. The first is that, on a fundamental level, the galaxy fluctuation field on small smoothing scales could become $\delta_g<-1$ which corresponds to a negative galaxy density. The second is that, from an observational point of view, this scheme  leaves the shape of the power spectrum unchanged while not all galaxy populations have the same observed power spectrum shape, although the differences are not large (e.g., \citet{peacock/dodds:1994,norberg/etal:2001}). Many different and more complicated biasing schemes have been introduced in the literature. For our purposes it is important to note that the bias of a sample of galaxies  depends on the sample selection criteria and on the weighting scheme used in the analysis. Thus different surveys will have different biases, and care must be taken when comparing the different galaxy power spectra.

There are several indications that large-scales galaxy bias is scale independent on large scales (e.g., \citet{hoekstra/etal:2002c,verde/etal:2002}). This justifies adopting equation \ref{eq:biasedpowersp}.  For the 2dFGRS, the bias of galaxies has been measured by Verde et al. (2002), by using higher-order correlations of the galaxy fluctuation field. They assume a generalization of the simple linear biasing scheme, $\delta_g=b_1\delta+b_2/2\delta^2$. They find no evidence for scale-dependent bias at least on linear and mildly non-linear scales (i.e. $k<0.4$ $h$ Mpc$^{-1}$) and $b_2$ consistent with $0$. This finding further supports the use of equation \ref{eq:biasedpowersp}. In particular, they find $b_1=1.06\pm 0.11$. In our analysis we will assume linear biasing.

The \citet{verde/etal:2002} bias measurement has to be interpreted with care. It applies to 2dFGRS galaxies weighted with a modification of the \citet{feldman/kaiser/peacock:1994} weighting scheme as described in Percival et al. (2001). It is important to note that, close to the observer, dim galaxies are included in the survey;  the galaxy density is high, but a  small volume of the sky is covered. On the other hand, far away from the observer, only very bright galaxies are included in the survey; a large volume is probed but the galaxy density is low. As a consequence, clustering of dim galaxies in a small volume close to the observer contain most of the  signal for the power spectrum at small scales. While rare, bright galaxies in a large volume enclose most of the information about the power spectrum on large scales. An ``optimal'' weighting scheme would thus weight dim galaxies on small scales and bright galaxies on large scales. This weighting scheme is, unfortunately, biased. Bright galaxies are more strongly clustered (i.e. more biased) than dim ones. This effect is known as ``luminosity bias''. The power spectrum recovered from such a weighting scheme will have optimal error bars, but will exhibit scale-dependent bias. The weighting scheme used in Percival et al. (2001) is not optimal, but is virtually unaffected by luminosity bias (Percival, Verde \& Peacock  2003). The power spectrum so obtained is that of 2 $L_*$ galaxies on virtually all scales, and the effective redshift for the power spectrum is $z_{\rm eff}=0.17$, slightly larger than the effective redshift of the survey as defined by the selection function \citep{percival/etal:2001,peacock/etal:2001}.

The final complication is that galaxy catalogs use the redshift as the third spatial coordinate. In a perfectly homogeneous Friedman universe, redshift would be an accurate distance indicator. Inhomogeneities, though, perturb the Hubble flow and introduce peculiar velocities. As \citet{kaiser:1987} emphasized, the peculiar velocities distort the clustering pattern not only on small scales where virialized objects produce ``Fingers-of-God'', but also on large scales where coherent flows produce large scale distortion components.

On large (linear) scales the redshift-space effect on an individual Fourier component of the density fluctuation field $\delta_{\bf k}$ can be modeled by,
\be
\delta_{\bf k}\longrightarrow \delta_{\bf k}^s=\delta_{\bf k}(1+\beta \mu^2)\;,
\ee
where the superscript $s$ refers to the quantity in redshift space and $\mu$ is the cosine of the angle between the $k$-vector and the line of sight. The Kaiser factor, $\beta$, is the linear redshift space distortion parameter.
One defines $\beta=f/b$, where
$f=d\ln \delta /d \ln a$, with $\delta=\delta\rho/\rho$ and $a=(1+z)^{-1}$;
$b$ is the linear bias parameter. The expression for $f(z)$ is a known function of $\Omega_m$, $\Lambda$ and $z$ \citep{lahav/etal:1991},
\begin{equation}
f(\Omega_m,\Lambda,z)=X^{-1}\left[\frac{\Lambda}{(1+z)^2}-\frac{\Omega_m}{2}(1+z)\right]-1+(1+z)^{-1}X^{-3/2}\left[\int_0^{\frac{1}{(1+z)}}X^{-2/3}da\right]^{-1}\;,
\label{eq:f}
\end{equation}
where $X=1+\Omega_mz+\Lambda(a^2-1)$, and can be approximated by\footnote{In our analysis we use the exact expression for $\beta$ as in equation(\ref{eq:f}).} $\beta \simeq \Omega_m^{0.6}/b$. The analysis of the 2dFGRS \citep{peacock/etal:2001,percival/etal:2001} constrains $f$ at the effective redshift of the survey. The effective redshift of the survey depends on the galaxy weighting scheme adopted to compute the power spectrum for the above work ($z_{\rm eff} \sim 0.17$). This peculiar velocity infall causes the overdensity to appear squashed along the line of sight. The net effect on the angle-averaged power spectrum in the small angle approximation is 
\be
P^s(k)=P(k)\left(1+\frac{2}{3}\beta+\frac{1}{5}\beta^2\right).
\ee
Thus on large scales the redshift space distortions boost the power spectrum if $\beta >0$.

On smaller scales, virialized motions produce a radial smearing and the associated ``Fingers-of-God'' effect contaminates the wavelengths we are interested in. This is difficult to treat exactly, but as it is a smearing effect, it produces a mild damping of the power, acting in the opposite direction to the large-scale boosting by the Kaiser effect (see for example \citet{matsubara:1994}). On these scales, the redshift space correlation function is well modeled as a convolution of the real space isotropic correlation function with some distribution function for the line of sight velocities (e.g. \citet{davis/peebles:1983,cole/fisher/weinberg:1994,fisher:1995}).
Since the convolution in real space is equivalent to multiplication in Fourier space, the redshift space power spectrum on small scales is multiplied by the square of the Fourier transform of the velocity distribution function (e.g., \citet{peacock/dodds:1994}), 
\be
P^s(k,\mu)=P(k)(1+\beta\mu^2)^2D(k\sigma_p\mu)\;,
\label{eq:powerspredshiftmu}
\ee
where $\sigma_p$ denotes the line-of-sight pairwise velocity dispersion.  If the pairwise velocity distribution is taken to be an exponential (e.g. \citet{ballinger/heavens/taylor:1995,ballinger/peacock/heavens:1996,hatton/cole:1998}) which seems to be supported by simulations (e.g., \citet{zurek/etal:1994}) and observations (e.g., \citet{marzke/etal:1995}), then the damping factor is a
Lorentzian (see also \citet{kang/etal:2002}),
\be
D(k\sigma_p\mu)=\frac{1}{1+(k^2\sigma_p^2\mu^2)/2}\;.
\label{eq:dampingexpon}
\ee  
We adopt this functional form as it is used by Peacock et al. (2001) in determining the redshift space distortion parameters $\beta$ and $\sigma_p$ from the
2dFGRS.
The overall effect for the power spectrum in a thin shell in $k$-space is given
by
\be
P^s(k)=\left[4\frac{(\sigma_p^2k^2-\beta)\beta}{\sigma_p^4k^4}+\frac{2\beta^2}{3\sigma_p^2k^2}+\frac{\sqrt{2}(k^2\sigma_p^2-2\beta)^2\mbox{arctan}(k\sigma_p/\sqrt{2})}{k^5\sigma_p^5}\right]P(k)
\label{eq:powerspectrumredshift}
\ee
obtained by averaging over $\mu$ in equation (\ref{eq:powerspredshiftmu}) with the damping factor given by equation (\ref{eq:dampingexpon}).
Figure \ref{fig:powerspredsh} show the effect of redshift space distortions (equation \ref{eq:powerspectrumredshift}) on the scales of interest.

This model is simplistic for several reasons. The most important is that, because of the complicated geometry of the survey, the simple angle average operation performed to obtain equation (\ref{eq:powerspectrumredshift}) might not be strictly correct. Also, equation (\ref{eq:powerspectrumredshift}) is obtained in the plane-parallel (also known as small-angle) approximation (i.e. as if the lines of sight to different galaxies on the sky were parallel).

We have performed extensive testing of equation (\ref{eq:powerspectrumredshift}) using mock 2dFGRS catalogs obtained from the Hubble volume simulation. We find  that the simulations redshift-space power spectrum is  consistent, given the errors, with equation (\ref{eq:powerspectrumredshift}) where $P(k)$ is the  simulations real-space power spectrum up to $k<0.4$ $h$ Mpc$^{-1}$, even for the complicated geometry of the 2dFGRS. This means that up to $k\sim 0.4$ the systematics introduced by eq. (\ref{eq:powerspectrumredshift}) are smaller than the statistical errors; in the analysis we use only $k\lesssim 0.15$.''
However, the value for $\beta$ in equation  (\ref{eq:powerspectrumredshift}) needs to be calibrated on Monte Carlo realizations of the survey. We find that $\beta^{\rm eff }=0.85 \beta$. We have verified that our results for the cosmological parameters are insensitive to the exact choice of the correction factor.  
Peacock et al. (2001) measured the parameters $\beta$ and $\sigma_p$ and their joint probability distribution from the survey obtaining  $\beta=0.43$ and $\sigma_p=385$ km s$^{-1}$. 
This measurement has been obtained by using the full angular dependence of the power spectrum and therefore recovers directly $\beta$ and not $\beta^{\rm eff}$. \citet{hawkins/etal:2002} measured these parameters from a larger sample than the one from Peacock et al. (2001), obtaining a slightly different result.
This is mostly due to a shift in the recovered value for $\sigma_p$. Since most of the galaxies in the Hawkings et al. (2002) sample are in the Peacock et al. (2001) sample, we conservatively extend our error-bars on $\beta$ and  $\sigma$  by 10\% and 30\% respectively, to include the new value within the 1-$\sigma$ marginalized confidence contour, and to include a possible error in the determination of $\beta^{\rm eff}$. Figure \ref{fig:powerspredsh} illustrates the importance of including all the above effects in our analysis.

In our analysis we consider data in the $k$ range $0.02< k <0.2$ $h$ Mpc$^{-1}$. On large scales the limit is set by the accuracy of the window function model; on small scales the limit is set by where the covariance matrix has been extensively tested. In this regime we also have a weak dependence on the velocity dispersion parameter $\sigma_p$, the parameter with the largest systematic uncertainty.

\subsubsection{Motivation for this Modeling}
\label{sec:motivation}

The motivation behind the complicated modeling of \S \ref{sec:surveygeom}-\ref{sec:galdistr} is to be able to infer the amplitude of the matter power spectrum from the observed galaxy clustering properties. 

\begin{figure}
\epsscale{0.8}
\plotone{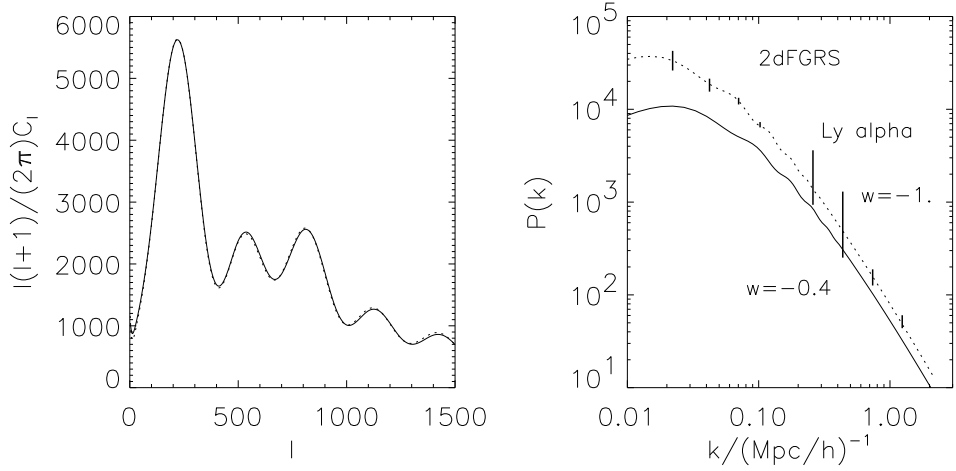}
\caption{\small Two cosmological models: $\omega_b=0.0235$, $\omega_m=0.143$, $n_s=0.978$ $\tau=0.11$, $w=-0.426$, $h=0.53$  (solid line) $\omega_b=0.0254$,  $\omega_m=0.137$, $n_s=1.024$, $\tau=0.08$  $w=-1$,  $h=0.77$ (dotted line). The two models are indistinguishable  within current error-bars from the CMB angular power spectrum (left panel, units for the power spectrum are $\mu$K$^2$). However they can easily be distinguished if we can relate the observed power spectrum to the underlying matter power spectrum (right panel, units for the power spectrum are (Mpc $h^{-1}$)$^3$). The error bars on the solid line are examples of the size of the 2dFGRS and \lya power spectra statistical error-bars for one data point at different scales. There are 4 error bars for 2dFGRS and 4 for \lya.}
\label{fig:wdegen}
\end{figure}

\begin{figure}
\epsscale{0.8}
\plotone{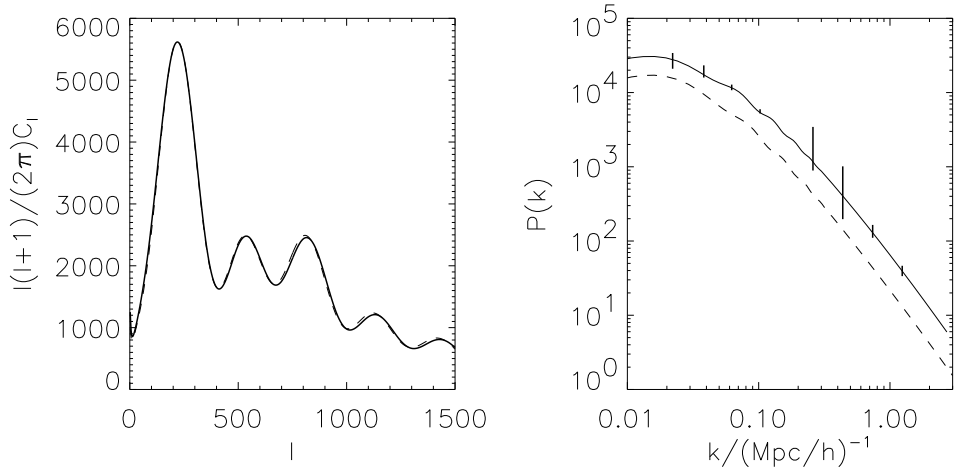}

\caption{\small Two cosmological models: $\Omega_m=0.26$, $\omega_b=0.02319$, $\tau=0.12$, $n_s=0.953$, $\omega_{\nu}=0$, $h=0.714$ (solid line) and $\Omega_m=0.26$, $\omega_b=0.02319$, $\tau=0.12$, $n_s=0.953$, $\omega_{\nu}=0.02$, $h=0.6$ (dashed line). As before the two models are virtually indistinguishable from the CMB angular power spectrum (left panel, units for the power spectrum are $\mu$K$^2$), but they can easily be distinguished if the matter power spectrum amplitude is known (right panel , units for the power spectrum are (Mpc $h^{-1}$)$^3$). The error bars on the solid line are examples of the size of the 2dFGRS and \lya power spectra statistical error-bars for one data point. There are 4 error bars for 2dFGRS and 4 for \lya. }
\label{fig:nudegen}
\end{figure}

Figures \ref{fig:wdegen} and \ref{fig:nudegen}  illustrate how the modeling of \S \ref{sec:surveygeom}-\ref{sec:galdistr} helps in breaking degeneracies among cosmological parameters. For illustration, we consider two cases below; the degeneracy of the dark energy equation of state, $w$, \citep{huey/etal:1999} with $\Omega_b$ and $n_s$ and the $\omega_{\nu}-h$ degeneracy, where $\omega_{\nu}=\Omega_{\nu}h^2$.  

Figure \ref{fig:wdegen} shows two models that are virtually indistinguishable with CMB data, but which predict different amplitudes for the matter power spectra at $z\sim 0$. This is because the linear growth factor and the shape parameter $\Gamma$ are different for the two cases. The two models differ in the values of $\omega_b$, $n_s$ and $w$. The solid line is a model with $w=-0.4$ while the dotted line is a model with $w=-1$.
 
In Figure \ref{fig:nudegen} we show two sets of cosmological parameters that differ only in the values of the neutrino mass and the Hubble constant. These two models are virtually indistinguishable with CMB observations. 
But the matter power spectrum in the two cases is different in shape and amplitude. Since redshift-space distortions and window function affect the power spectrum shape, extra information about cosmological parameters is encoded in its amplitude. By using this information, \citet{spergel/etal:2003} obtain a cosmological upper bound on the neutrino mass that is $\sim 4$ times better than current cosmological constraints \citep{Elgaroy/etal:2002}. 

For completeness, we have shown the power spectrum also for scales probed by the \lya forest (see \S \ref{sec:Lya}). The error bars in Figures \ref{fig:wdegen} and \ref{fig:nudegen} are examples of the size of  the 2dFGRS and \lya power spectra statistical uncertainties in one data-point, showing that the two models can be distinguished if the observed power spectrum can be related to the linear matter power spectrum without introducing large additional uncertainties.

\subsubsection{Practical Approach}
The procedure we adopt in order to compare the observed galaxy power spectrum with the theory predictions is outlined below (the published 2dFGRS galaxy power spectrum has been already corrected for shot noise).
For a given set of cosmological parameters  and a pair-wise velocity dispersion parameter we proceed as follows: 
\begin{itemize}
\item[1)] The MCMC selects a set of cosmological parameters and values for $\beta$ and $\sigma_p$. {\sf CMBFAST} computes the theoretical linear matter power spectrum at $z=0$. 
\item[2)] We evolve the theoretical linear matter power spectrum  to obtain the non-linear matter power spectrum at the effective redshift of the survey, following the prescription of \citet{ma/etal:1999}.
\item[3)] We then obtain the redshift-space power spectrum for the mass by using equation \ref{eq:powerspectrumredshift} with $\beta^{\rm eff}$ calibrated from Monte Carlo realizations of the catalog.
\item[4)] The bias is computed from $\beta$ and $\Omega_m$ using equation (\ref{eq:f}). The galaxy power spectrum is obtained by correcting for bias, equation \ref{eq:biasedpowersp}. 
\item[5)] The resulting power spectrum is convolved with the galaxy window function. We use the routine provided on the 2dFGRS web site to perform this numerically. This is the power spectrum that can be compared with the quantity measured from a galaxy survey.
\item[6)] We can now evaluate the likelihood using the full covariance matrix as provided by the 2dFGRS team. We approximate the likelihood to be Gaussian as it was done by the team. In principle this is not strictly correct since in the linear regime the power spectrum is an exponential distribution and in the non-linear regime the distribution has contributions from higher-orders correlations. However, due to the size of the survey we are considering, the central limit theorem ensures that the likelihood is well described by the Gaussian approximation (e.g., \citet{scoccimarro/zaldarriaga/hui:1999}). Moreover, the covariance matrix for the 2dFGRS power spectrum has been computed by the 2dFGRS team under the assumption that the likelihood is
Gaussian.
\end{itemize}

We assume that the likelihood for the bias parameter is Gaussian, centered on $b=1.04$ with dispersion $\sigma_b=0.11$. This is a conservative overestimate of the error on the bias parameter, as noted in Verde et al. (2002). The determination of $b$ is correlated with  $\beta$ and $\sigma_p$ and the error quoted has already been marginalized over the uncertainties in these two parameters.  In practice, for each step in the Markov chain we compute the likelihood according to items 1 through 6 above. The bias parameter is determined once $\beta$, $\sigma_p$ and the other cosmological parameters are chosen. We then multiply the likelihood by the joint likelihood for $\beta$ and $\sigma_p$, as in Figure 4 of \citet{peacock/etal:2001}, and by the likelihood for the bias parameter. In effect, we use the determination of $\beta$, $\sigma_p$, and $b$ as priors.  By multiplying the likelihood we assume that the {\it measurements} of the redshift space distortion parameters, bias, and the 2dFGRS power spectrum are independent. We justify this assumption below.

The parameters needed to map the real-space non-linear matter power spectrum onto the redshift-space galaxy spectrum are: $\beta$, $\sigma_p$  and $b$.
These three parameters are not independent, not only is $\beta \propto 1/b$ but, more importantly, the three parameters are measured from the same catalog which we are using to constrain  other cosmological parameters. However, the information we use to constrain cosmological parameters is all encoded in the shape and amplitude of the angle-averaged power spectrum. The information used to measure $\beta$ and $\sigma_p$ is all encoded in the dependence of the Fourier coefficients (i.e., of the power spectrum) on the angle from the line of sight. Thus we can treat the determinations of $\beta$ and $\sigma_p$ as independent from the likelihood for cosmological parameters. The analysis of \citet{verde/etal:2002} to measure the bias parameter from the 2dFGRS uses both information about the amplitude of the Fourier coefficients and their angular dependence. This dependence, however, is not that introduced by redshift-space-distortions, but is the configuration dependence of the bispectrum.  
Thus, in principle we should not treat this measurement as completely independent. However, most of the signal for the bias measurement comes from the $k$-range of $0.2 < k < 0.4$ $h$ Mpc$^{-1}$ while the signal for the present analysis comes from $k<0.2$ $h$ Mpc$^{-1}$. Note that the configuration dependence of the bispectrum is largely independent of cosmology. This allows us to easily include a prior for the bias parameter in the analysis.  
 
\section{LYMAN $\alpha$  FOREST DATA}
\label{sec:Lya}

The \lya forest traces the fluctuations in the neutral gas density along the line of sight to distant quasars.  Since most of this absorption is produced by low density unshocked gas in the voids or in mildly overdense regions that are thought to be in ionization equilibrium, this gas is assumed to be an accurate tracer of the large-scale distribution of dark matter. In this epoch and on these scales the clustering of dark matter is still in the linear regime. 

Since the \lya forest observations are probing the distribution of matter at $z \sim 3$, they are an important complement to the CMB data and the galaxy surveys data.  Because of their importance, there has been extensive numerical and observational work testing the notion that they trace the large-scale structure.
In our analyses, we find that the addition of Lyman $\alpha$ forest data appear to confirm trends seen in other data sets and tightens cosmological constraints. However, more observational and theoretical work is still needed to confirm the validity of the emerging consensus that the \lya forest data traces the LSS.

Recent papers use two different approaches for analysis of the \lya forest power spectrum data. \citet{mcdonald/etal:2000} and \citet{zaldarriaga/hui/tegmark:2001} directly compare the observed transmission spectra to the predictions from cosmological models.  We follow the approach of \citet{croft/etal:2002} and \citet{gnedin/hamilton:2002}(GH) who use an analytical fitting function to recover the matter power spectrum from the transmission spectrum\footnote{After the present paper was submitted, a preprint appeared (Seljak et al. 2003) claiming that the treatement of GH and \citet{croft/etal:2002} significantly underestimate the errors. Given the importance of this data set to tighten cosmological constraints,  the \lya  forest community should  reach a consensus on the interpretation of these observations  and on the level of systematic contamination}. 

GH factorize the linear power spectrum into four terms,
\begin{equation}
P_L(k)=P^{\rm fact}(k)Q_{\Omega}Q_{T}Q_{\tau}\;,
\label{eqlyalphaQ}
\end{equation}
where $P^{\rm fact}(k)$ is a quantity that is independent of modeling and is  is almost directly measured. The other parameters convert this quantity into the linear matter power spectrum and encode the  dependence on cosmology and the modeling of the inter galactic medium (IGM). In our treatment, we use the values of $P^{\rm obs}(k)$ (the estimator from \lya forest  observations of $P^{\rm fact}$) from GH and their parameterization in terms of equation (\ref{eqlyalphaQ}) because it allows us to explicitly include the dependence of the recovered linear matter power spectrum on the cosmological parameters. $Q_{\Omega}$ encodes the dependence of the recovered linear power spectrum  on the matter density parameter at $z=2.72$. For $Q_{\Omega}$ we use the  GH ansatz of,
\begin{equation}
Q_{\Omega}=\left(\frac{2.4}{1+\Omega_m^{0.6}1.4}\right)^2\;.
\end{equation}
$Q_T=20,000$ ${\rm K}/T_0$ ($T_0 \sim 20,000$ K) parameterizes the dependence on the mean temperature of the IGM, $Q_{\tau}\sim 1.11$ parameterizes the dependence on the assumed mean optical depth.
In addition to the statistical errors, GH quote a systematic uncertainty that we add to the statistical one. Finally, the uncertainties in $Q_{\Omega}, Q_{T}$ and $Q_{\tau}$ contribute to the overall normalization uncertainty.  We use the \citet{croft/etal:2002} prescription to parameterize this uncertainty as $\ln{\cal  P}(A)=-1/2(A-1)^2/\sigma_{Ly_{\alpha}}^2$ where if $A\le 1$ then $\sigma_{Ly_{\alpha}}=0.25$ while if $A > 1$, $\sigma_{Ly_{\alpha}}=0.29$.

 N-body simulations are used to convert the flux power spectrum to the dark matter power spectrum and calibrate the form of equation (\ref {eqlyalphaQ}). The two different groups, GH and \citet{croft/etal:2002}, have done this independently. The resulting power spectra agree well within the 1-$\sigma$ errors for all data points except the last three. We thus increase the 1$\sigma$ uncertainties on the last three data points to make the two determinations of $P_L(k)$ consistent and use this as a rough measure of the intrinsic systematic uncertainties in the \lya data.

GH point out that the correlation in flux measured from the \lya forest samples power over a finite band of wavenumbers. The effective band-power windows are rather broad due to the peculiar velocities that smear power on scales comparable to the 1-d velocity dispersion.  Thus the recovered linear power spectrum is effectively smoothed with an window that becomes broader at smaller scales. In principle, the resulting covariance between estimates of power at different $k$ needs to be taken into account to do a full likelihood analysis to extract cosmological parameters. However, the full covariance matrix is not available. Since the \lya data are such a  powerful tool 
we  just perform a simple $\chi^2$ fit and caution the reader that interpreting the reduced $\chi^2$ as a measure of goodness of fit for this data set is not meaningful since the data are strongly correlated.

To marginalize over the overall normalization uncertainty, we take advantage of the MCMC approach. In principle we could marginalize over it analytically, as we do for the calibration uncertainty. Instead, at each step of the chain we compute the best fit amplitude $\bar{A}$ as done for point sources \citep{hinshaw/etal:2003},
\be
 \bar{A}=\left[\sum_k (P^{\rm obs}(k)P_L(k)/\sigma_k^2)\right]\left[\sum_k(P^{\rm obs}(k)^2/\sigma_k^2)\right]^{-1}\;.
\ee
The likelihood for the \lya  data for the model is given by $\ln {\cal L}_{Ly_{\alpha}}=\ln {\cal L}(P^{\rm obs}|\bar{A},P_{L})+\ln{\cal P}(\bar{A})$. The marginalization is then automatically obtained from the MCMC output. In other words, the analytic marginalization computes $\int {\cal P}({\rm data}|{\rm model}){\cal P}(A) dA$ while we compute an estimator of this given by $\int {\cal P}({\rm data}|{\rm model}){\cal P}(\bar{A}) d\bar{A}$.

\section{Conclusions}

In this paper, we have presented the basic formalism that we use for our likelihood analysis. This paper shows the  final step on the path from time-ordered data to cosmological parameters.  It provides the framework for the analysis of cosmological parameters and  their implications for cosmology.

The unprecedented quality of the {\sl WMAP} data and the tight constraints on cosmological parameters that are derived require a rigorous analysis so that the approximations made in the modeling  do not propagate into significant  biases and systematic errors.  We have derived an approximation to the exact likelihood function for the ${\cal C}_{\ell}$ which is accurate to better than 0.1\%, and we have carefully  calibrated the temperature power spectrum covariance matrix with Monte Carlo simulations. This enables us to use the effective chi-squared per degree of freedom as a tool to test whether or not a model is an acceptable fit to the data.

We implement our likelihood analysis with the MCMC.  We have concentrated on the issue of convergence and mixing, emphasizing how important these issues are in recovering cosmological parameters values and their confidence levels from the MCMC output.

To the {\sl WMAP} data-sets (TT and TE angular power spectra) we have added the CBI and ACBAR measurement of the CMB on smaller angular scales, the 2dFGRS galaxy power spectrum at $z\sim 0$, and the \lya forest matter power spectrum at $z\sim 3$. These external data sets significantly enhance the scientific value of the {\sl WMAP} measurement, by allowing us to break 
parameter degeneracies.
While the underlying physics for these data sets is much more complicated and less well understood than for {\sl WMAP} data, and systematic and instrumental effects are much more important, we feel we have made a significant step towards improving the rigor of the analysis of these data sets.
We have included a detailed modeling of galaxy bias, redshift distortions and the non-linear growth of structure. We also include known (as to the present day) systematic and statistical  uncertainties  intrinsic to these other data sets.

\section*{Acknowledgments}

We thank Bill Holzapfel for invaluable discussions about the ACBAR data. We thank the 2dFGRS team for giving us access to the Monte Carlo realizations of the 2dFGRS. The mock catalogs of the 2dFGRS were constructed at the Institute for Computational Cosmology at Durham. We thank Will Percival for discussions and for providing us with the covariance matrix of the 2dFGRS power spectrum. LV is supported by NASA through Chandra Fellowship PF2-30022 issued by the Chandra X-ray Observatory center, which is operated by the Smithsonian Astrophysical Observatory for and on behalf of  NASA under contract NAS8-39073. LV also acknowledges Rutgers University for support during the initial stages of this work. HVP is supported by a Dodds fellowship granted by Princeton University.
The {\WMAP} mission is made possible by the support of the Office of Space 
Sciences at NASA Headquarters and by the hard and capable work of scores of 
scientists, engineers, technicians, machinists, data analysts, 
budget analysts, managers, administrative staff, and reviewers.


\end{document}